\documentclass{emulateapj}

\usepackage{natbib}
\usepackage{graphicx}

\newcommand{\teff}{\textit{T}\textsubscript{eff}}
\newcommand{\zmax}{$Z_{\textrm{max}}$}
\newcommand{\rearth}{$R_{\oplus}$}
\newcommand{\f}{$f$}
\newcommand{\fone}{$f_1$}
\newcommand{\ftwo}{$f_2$}
\newcommand{\threshold}{$t$}

\newcommand{\logg}{log \textit{g}}
\usepackage{mathrsfs}
\usepackage{comment}
\usepackage{gensymb} 

\usepackage[T1]{fontenc}
\usepackage{ae,aecompl}
\usepackage{breakurl}

\usepackage{amsmath}
\usepackage{amsfonts}
\usepackage{amssymb}
\usepackage{comment}
\usepackage[dvipsnames]{xcolor}
\usepackage[figuresleft]{rotating}

\usepackage[%
        ]{hyperref}

\usepackage{xcolor}
\usepackage{ulem}   
\definecolor{kcolor}{rgb}{0.54, 0.17, 0.89}

\begin{document}


\title{An Increase in the Galactic Planet Host Fraction Fails to Reproduce the Galactic Height Trend in Planet Occurrence}

\author{Christopher~Lam\altaffilmark{1,2}, Sarah~Ballard\altaffilmark{2}, Sheila~Sagear\altaffilmark{2}, \& Kathryne~J.~Daniel\altaffilmark{3}}

\altaffiltext{1}{IPAC, California Institute of Technology, Pasadena, CA 91125, USA; lamchris@ipac.caltech.edu}
\altaffiltext{2}{University of Florida Department of Astronomy, 211 Bryant Space Science Center, Gainesville, FL 32611, USA}
\altaffiltext{3}{Department of Astronomy \& Steward Observatory, University of Arizona, 933 North Cherry Avenue, Tucson, AZ 85721, USA}






\begin{abstract}
 While stellar metallicity has long been known to correlate with planetary properties, the galactic metallicity gradient alone does not account for the observed strong trend in planet occurrence with Galactic height. In this study, we investigate the observable effect of a time-dependent planet occurrence rate upon a sample of stars selected uniformly from the \textit{Kepler} and K2 surveys. Using a novel planetary system population synthesis code, \texttt{psps}, we impose several prescriptions for a time-variable planet host fraction, \f, in which a primordial \fone\ either instantaneously or gradually increased to a present-day \ftwo. We then simulate the expected small planet occurrence rate around FGK dwarfs as a function of galactic height. Finally, we compare the modeled trends to the observed result from the missions themselves. We find that using a joint \textit{Kepler}-K2 sample with isochrone ages, an increase in \f\ is insufficient to reproduce the strength of the observed trend between occurrence and Galactic height. We show that not all of this is due to insufficient age precision: using a synthetic stellar population from the TRILEGAL framework, we show that even with very precise ages, we can rule out models of gradually increasing \f. We also derive the \textit{Kepler} occurrence-height relation and find that an increase in \f\ is better able to match this trend. An analysis using more precise ages and incorporating an evolving compact multi fraction could furnish a realistic relation in planet occurrence with Galactic height that matches the observed \textit{Kepler}-K2 trend.
\end{abstract}

\section{Introduction}
\label{sec:introduction}

The canonical story of planet formation is that of a localized, isolated process \citep{Armitage11, Williams11, Winn15}, largely independent of galactic-scale phenomena. Within this picture, to the extent that the galactic context matters, it is through the steady enrichment of metals in the interstellar medium (ISM; \cite{nielsen_planet_2023}). Stellar metallicity traces the metallicity and mass of the protoplanetary disk \citep{andrews_mass_2013}. This, in turn, is thought to determine planet radii and system architectures \citep{santos_metal-rich_2001, fischer_planet-metallicity_2005, Mordasini12, brewer_compact_2018, petigura_california-kepler_2018, boley_first_2024, bryan_friends_2024, buchhave_abundance_2012}. 

However, recent observational results challenge the notion that metallicity alone can explain variations in planet occurrence on galactic scales. The extent to which metallicity is itself deterministic, or whether it is a tracer for other processes that shape planet formation, is under active debate. Establishing the relationship between host star metallicity and planet occurrence is itself complicated. For example, metallicity and close binarity \citep{Kraus16, moe_impact_2021} both affect planet outcomes, but they are related to one another (see review by \citealt{Moe19}) and observationally entangled \citep{Furlan20}. This has led some studies to conclude that the masses of planets are determined by factors other than the availability of solids per se \citep{kutra_super-earths_2021}, instead regulated by an as-yet unknown process.  Stellar mass, effective temperature, and age are similar quantities in the planet formation story. They are related to planet outcomes (see e.g. \citealt{howard_planet_2012, mulders_increase_2015, he_architectures_2020-1, berger_gaiakepler_2020-1, sayeed_exoplanet_2025, yang_occurrence_2020}) but also related to one another. Disambiguating these effects poses a major challenge to understanding occurrence. For example, \cite{Fulton18} observed that among \textit{Kepler} stars, more massive stars are more metal-rich. However, this reflects a selection bias whereby more massive stars are likelier to be younger, and thus formed from more recent and enriched galactic material (see also \citealt{kutra_super-earths_2021}). On top of these quantities that trace galactic star formation history, planetary systems may themselves evolve over time, which can potentially masquerade as a metallicity-dependent effect. Planetary systems could potentially be self-disrupting \citep{pu_spacing_2015, lam_ages_2024}. Alternatively, they may change their configuration due to stellar flybys or galatic tides: these can either directly impact outer planets and potentially propagate inward \citep{zakamska_excitation_2004, rodet_correlation_2021, veras_exoplanets_2013}, or affect a binary companion to the host star, whose altered orbit might then disrupt planets \citep{kaib_planetary_2013, correa-otto_galactic_2017}.

The \textit{Gaia} mission \citep{Prusti16} has spurred a transformation in exoplanetary studies, driven by an improved understanding of host star properties and also their location and movement within the galaxy. Recent studies have begun placing the relationship between metallicity and planet demographics in the context of the host stars' position in the Milky Way and dynamical history \citep{winter_stellar_2020, kruijssen_bridging_2020, nielsen_planet_2023, kruijssen_not_2021, bashi_exoplanets_2022, yang_planets_2023}. Within this new galactic framework for exoplanets, it is necessary to fold together our understanding of exoplanet demographics with a picture of how stars orbit the galactic potential. These orbits can change significantly over the stellar lifetime. Changes occur as stars repeatedly experience gravitational interactions with inhomogeneities in the galactic disk, such as giant molecular clouds (GMCs) \citep{SpitzerSchwarzschild1951, Wielen1977, Lacey1984},
spiral arms \citep{BarbanisWoltjer1967, CarlbergSellwood1985,
MinchevQuillen2006}, and the bar \citep{SahaTsengTaam2010,
Grand2016}, torques from misaligned stellar and gas disks \citep{Roskar2010,Khachaturyants2022}, feedback-driven fluctuations in the potential \citep{El-Badry2016}, as well as effects from the cosmological environment such as satellite interactions and mergers \citep[e.g.,][]{QuinnHernquistFullagar1993,
Brook2004, VillalobosHelmi2008, Bird2012}.  Interactions with lumps in the disk mass distribution, such as GMCs, can convert in-plane motions to vertical excursions  from the plane of the disk \citep[e.g,][]{Lacey1984, CarlbergInnanen1987, JenkinsBinney1990, Sellwood2013}, and satellite bombardment can significantly increase the vertical velocity dispersion of the disk \citep{Bird2012}.  Additionally, there is growing evidence that stellar populations formed during the early stages of the disk (${\sim}8-4$~Gyr ago in the Milky Way) were born with higher vertical velocity dispersion than in the current epoch \citep[see][and references therein]{McCluskey2024,Bird2021,Bland-HawthornGerhard2016}. Broadly speaking, older stellar populations are kinematically warmer than younger populations, an expectation reinforced by findings directly connecting \textit{Gaia} kinematic data to stellar age (see e.g. \citealt{mackereth_dynamical_2019, wu_agemetallicity_2021, casagrande_measuring_2016, mccluskey_disc_2024, iorio_chemo-kinematics_2021, gallart_uncovering_2019, sagear_zoomies_2024}). 

Situated at a key junction in this landscape is a recent result by \citet{zink_scaling_2023} that found that planet occurrence is higher among stars with lower oscillation amplitude from the galactic midplane: it decreases by a factor of several between 100-1000 pc (dependent upon planet size). Given the known relation between stellar metallicity and planet outcomes, a natural explanation is that the negative metallicity gradient with galactic scale height is responsible. This gradient -- which itself has been observed as far back as \citet{1976A&A....48..301M} and continues to be further constrained today \citep{yan_chemical_2019, Hayden2020, carrillo_relationship_2023, lian_integrated_2023,Sun2025} -- is not sufficient to explain the decrease in planet occurrence with increasing height from the midplane of the Milky Way, at least for small planets. Using a calculated raw exoplanet occurrence among \textit{Kepler} Super-Earths and Sub-Neptunes (planets with radius <4 $R_{\oplus}$) with a period of 1-40 days, as well as \textit{Gaia} stellar galactic oscillation amplitudes, \cite{zink_scaling_2023} fit a power law to model the slope of this relation, finding that the reduction in planet occurrence over 1 kpc is significantly greater than the expected reduction from a solely metallicity-driven occurrence-scale height trend. There is some tension between this finding and other studies, such as \citet{teixeira_where_2025}, who concluded that the planet formation rate of Neptune-sized and smaller planets increases slightly and gradually with age, in a way that is adequately explained by galactic metallicity evolution. 

The authors of \citet{zink_scaling_2023} (hereafter Z23) comment that the oscillation amplitude (hereafter denoted as \zmax) could be a proxy for another parameter like stellar age; this is suggestive of a phenomenological experiment, in which we are agnostic about the driver of this occurrence-height trend but can at least narrow down suspects by constraining the possible timescales on which such a driver may operate. Our aim in this study is to test whether and how a fiducial boost in the planet occurrence rate sometime in our Galaxy's past could have produced such a relation among close-in small planets around Sun-like stars. 



In Section \ref{sec:methods}, we describe our stellar sample (Section \ref{sec:stellar-sample}), an idealized, independently synthesized stellar sample (Section \ref{subsec:trilegal_stellar_sample}), our planet host fraction evolution models (Section \ref{sec:models}), the way in which we draw our synthetic planetary system populations (Section \ref{sec:psps}), and our completeness calculations (Section \ref{sec:completeness}). In Section \ref{sec:results}, we present the detection yields for each model and highlight those that come close to matching the Z23 relation between planet occurrence and galactic scale height. In Section \ref{sec:discussion}, we discuss the physical processes that might be represented by these favored models. Finally, we conclude in Section \ref{sec:conclusion}.

\section{Methods}
\label{sec:methods}

We aim to test whether and how a time-dependent planet occurrence rate, when folded together with typical kinematic heating of stars over Gyr, can reproduce the Z23 finding. To this end, we simulate suites of synthetic planetary system populations based on a prescription for the planet host fraction, \f, in the Milky Way as a function of stellar age. These models vary \f\ over time in ways we describe in this section. With an additional prescription for the kinematic heating of stellar orbits in the galaxy, we can synthesize ``present-day" planet occurrence yields that should vary with the host stars' height above the galactic midplane. To be clear, the exact ``height" quantity employed by Z23 for the host stars is not the instantaneous position above the midplane, but rather \zmax, the maximum vertical oscillation amplitude above and below the galactic midplane. In this study, we use the inverse detection efficiency method (IDEM), in which the detection efficiency is pre-computed and then applied to a detected yield in order to estimate a ``true" yield (see Section \ref{sec:completeness}). 

To streamline and generalize this workflow across different occurrence scenarios, we developed a planetary system population synthesis code in \texttt{Python} called \texttt{psps}\footnote{https://github.com/exoclam/psps}. The code enables rapid generation of planetary systems drawn probabilistically from user-defined prescriptions for various exoplanet demographic attributes. In Section \ref{sec:stellar-sample}, we describe how we construct the stellar samples in this work. In Section \ref{sec:models}, we define the time-dependent occurrence models used to determine which stars host planets. We then populate those systems using \texttt{psps} (Section \ref{sec:psps}) and ``observe" them with geometric and instrumental completeness models that reflect \textit{Kepler}'s detection sensitivity (Section \ref{sec:completeness}).

\subsection{Stellar Sample} \label{sec:stellar-sample}
The Z23 result was derived from a joint accounting of \textit{Kepler} and K2 stars. As such, in order to reproduce their result, we must begin with a stellar sample that uniformly includes both surveys. We construct this sample in two ways. First, we populate our sample with actual \textit{Kepler} and K2 stars, using their isochrone ages and computing their $Z_{\textrm{max}}$ based on kinematic information from \textit{Gaia} and a model of the Galactic potential. This approach has the benefit of matching the true host star sample exactly. However, it inherits the significant observational uncertainties associated with stellar age measurements from isochrones. For this first sample, we use the uniform catalog from \cite{hardegree-ullman_scaling_2025}. We describe this sample in Section \ref{subsec:real}. 

Second, we generate a synthetic \textit{Kepler} and K2 stellar sample using TRILEGAL \citep{girardi_star_2005}, a population synthesis code widely used in exoplanet demographic studies (e.g., \citealt{tamburo_predicting_2023, morton_discerning_2011, muirhead_catalog_2018, bouma_ages_2024}). While the TRILEGAL sample is not drawn from the \textit{Kepler} and Ecliptic Plane Input Catalogs, it can be tuned to resemble them reasonably well \citep{van_saders_forward_2019}, and it offers the advantage of generating arbitrarily large samples with small stellar age uncertainties. This allows us to apply time-dependent planet occurrence models with greater clarity. We describe this synthetic sample in Section \ref{subsec:trilegal_stellar_sample}. 

\subsubsection{Real Kepler and K2 sample}
\label{subsec:real}
We begin by generating stellar samples from homogeneously derived \textit{Kepler} and K2 catalogs, culling them in much the same manner as Z23 before enriching them with isochrone ages and \zmax. We use the homogeneously derived joint \textit{Kepler}-K2 catalog from \citet{hardegree-ullman_scaling_2025} (hereafter HU25), which already precludes likely astrometric binaries via a RUWE cut at 1.4 and noisy stars via CDPP cuts (only \textit{Kepler} stars with CDPP$_{7.5hr}$ < 1000 ppm and K2 stars with CDPP$_{8hr}$ < 1200 ppm are kept). From this catalog, we replicate the cuts made by Z23 and retain only stars that satisfy the following conditions:
\begin{enumerate}
    \item 4000 K $\leq$ \teff\ $\leq$ 6500 K
    \item \logg\ $\geq$ $\logg_{H16}$
\end{enumerate}
where H16 is the threshold from the heuristic devised by \cite{huber_k2_2016}:
\begin{equation} \label{eq:giant-cut}
    \frac{arctan(\frac{6300 K-\teff}{67.172 K})}{4.671} + 3.876.
\end{equation}

For \teff, \logg, and kinematic information, we use DR3 parameters from HU25 for both \textit{Kepler} and K2 samples. For the \textit{Kepler} and K2 isochrone ages, we draw from the \textit{Kepler}-\textit{Gaia} cross-match by \cite{berger_gaiakepler_2020} (hereafter B20) and the K2 portion of the \textit{Kepler}-K2-TESS-\textit{Gaia} cross-match by \cite{berger_gaiakeplertess-host_2025} (hereafter B25), respectively. From B20 and B25, we first remove stars with uninformative isochrone age posteriors (`unReAgeFlag' in B20 and `$f_{Age}$' in B25) and whose best-fit isochrone ages are less than 0.5 Gyr or greater than 8 Gyr. Next, we apply the same \teff\ and \logg\ cuts to B20 and B25 as we did to HU25 (and as did in Z23). We then cross-match the HU25 \textit{Kepler} catalog with B20 and the HU25 K2 catalog with the K2 subset of B25. This data preparation step introduces two possible sources of bias into our final sample: 1) the B25 sample contains only planet hosts; and 2) stellar ages from B20 are derived from \textit{Gaia} DR2 instead of DR3. 

\begin{figure}  
\includegraphics[width=.45\textwidth]{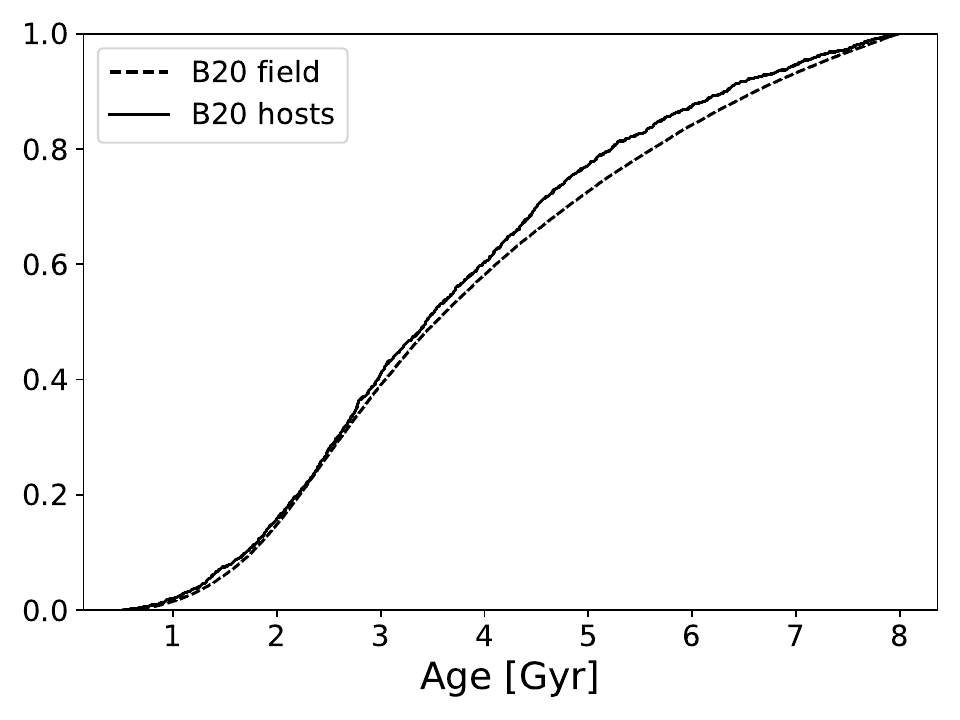}
\caption{Cumulative distribution functions of \textit{Gaia} DR2 isochrone ages for the B20 sample, split between hosts (solid) and non-hosts (dashed). A KS test shows that these age distributions are very likely drawn from essentially the same population, satisfying our check that the use of purely host ages from B25 does not introduce a selection bias.} 
\label{fig:b20-hosts-vs-field}
\end{figure}

To address the first potential issue, in Figure \ref{fig:b20-hosts-vs-field}, we verify that the age distributions of \textit{Kepler} stars are essentially the same between hosts and non-hosts. Comparing the cumulative distribution functions (CDFs) of the B20 host ages and the B20 non-host (also known as ``field star") ages, we find a Kolmogorov-Smirnov (KS) test statistic of 0.05 with a p-value of 7e-4, which indicates a very high probability that the two sets of ages are drawn from approximately the same population. We deem this result sufficient to move forward using B25 for the K2 ages.

\begin{figure*}  
\includegraphics[width=.45\textwidth]{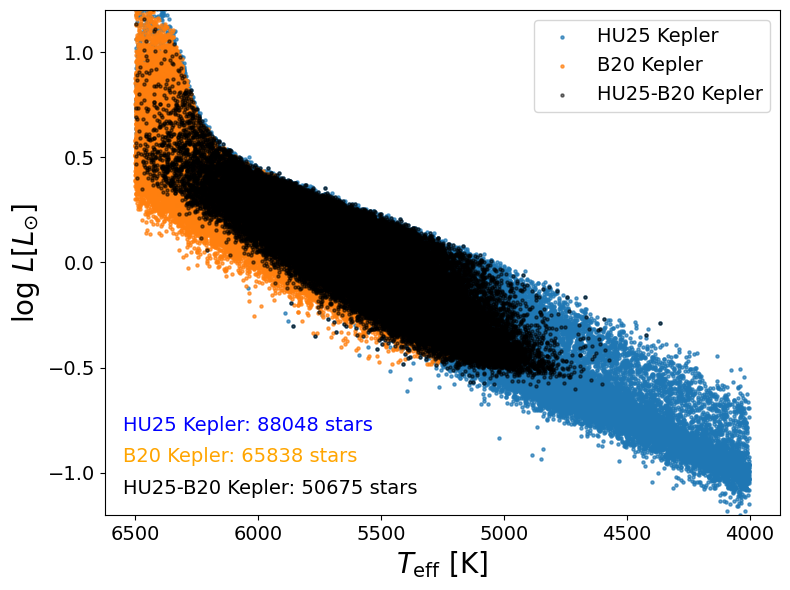}
\includegraphics[width=.45\textwidth]{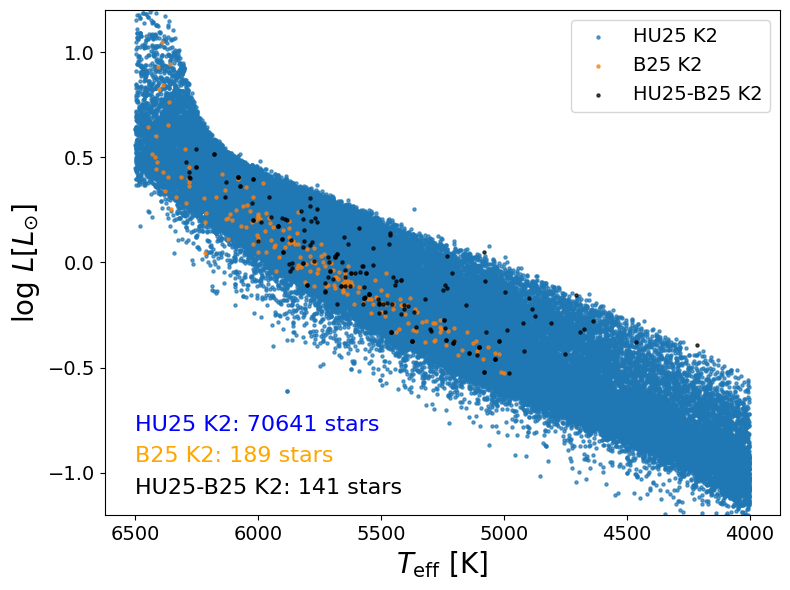}\\
\includegraphics[width=.45\textwidth]{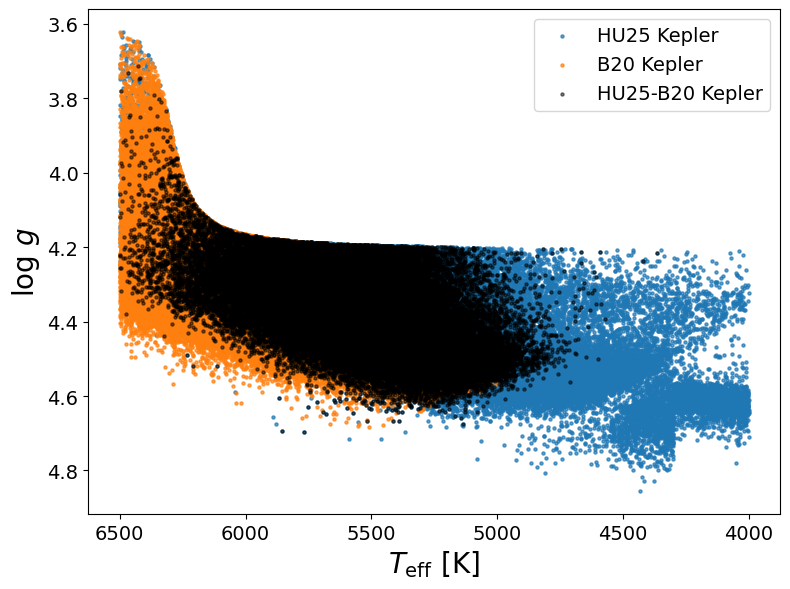}
\includegraphics[width=.45\textwidth]{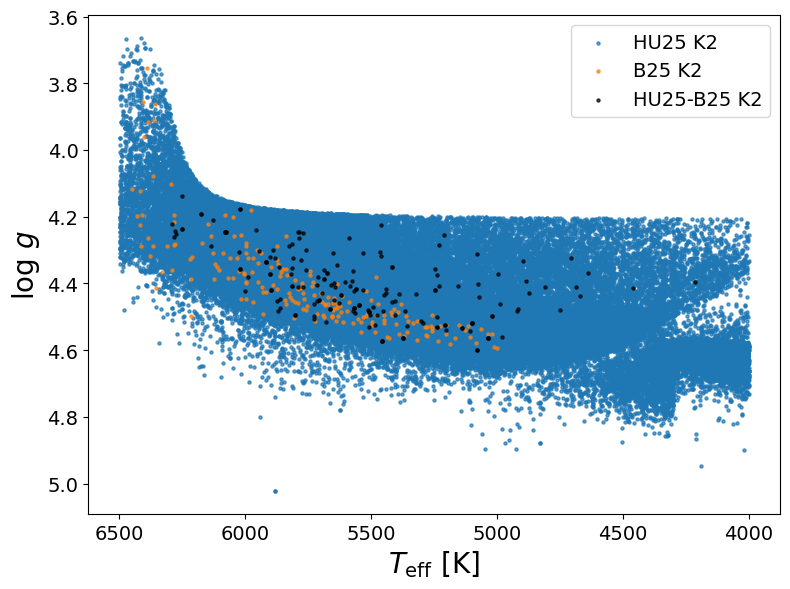}
\caption{\textbf{Top left: }HR diagram of the HU25 \textit{Kepler}, B20, and HU25-B20 cross-matched samples after culling by \teff, \logg, [Fe/H], and age (for B20). \textbf{Top right: } Same as top, but for the culled HU25 and B25 K2 samples and their cross-match. \textbf{Bottom left: } Kiel diagram of HU25 \textit{Kepler}, B20, and HU25-B20. \textbf{Bottom right: } Kiel diagram of HU25 K2, B25 K2, and HU25-B25 K2.} 
\label{fig:hr-hu25-b20-b25}
\end{figure*}

To address the second potential issue, we visually inspect the component and cross-matched catalogs' Hertzsprung-Russell (HR) and Kiel diagrams to verify that they occupy the same regions of relevant parameter space. The left panels of Figure \ref{fig:hr-hu25-b20-b25} show that while B20 does not extend to the cooler stars contained in HU25, it more importantly does not leave the footprint of the HU25 sample in \teff, \logg, and luminosity space, indicating that their cross-match does not introduce a bias in stellar temperature, surface gravity, or luminosity. This is less of a concern for K2, since B25 and HU25 both use \textit{Gaia} DR3, but we can still confirm in the right panels of Figure \ref{fig:hr-hu25-b20-b25} that B25 and HU25 are self-consistent in their HR and Kiel diagrams. The HR diagrams show the number of stars in each catalog after the \teff\ and \logg\ (and, for B20 and B25, age) cuts. These numbers, as well as the number of stars remaining at each intermediate parameter space cut, are also listed in Table \ref{tab:stellar-cuts}. 

\begin{table}[]
    \centering
    \caption{Number of stars after each stage of data preparation}
    \begin{tabular}{c|c}
       Data Prep Stage & Stars Remaining \\
       \hline
       \multicolumn{2}{c}{\textit{Kepler}} \\
       \hline
       HU25 initial & 94422 \\
       HU25 \teff\ cut & 88048 \\
       HU25 \logg\ cut & 88048 \\
       \hline
       B20 initial & 186301 \\
       B20 unreliable age cut & 150560 \\
       B20 \teff\ cut & 136244 \\
       B20 \logg\ cut & 81158 \\
       B20 age cut & 65838 \\
       \hline 
       HU25-B20 cross-match & 50675 \\
       HU25-B20 with full astrometry & 22043 \\
       \hline
       \multicolumn{2}{c}{K2} \\
       \hline
       HU25 initial & 99142 \\
       HU25 \teff\ cut & 70641 \\
       HU25 \logg\ cut & 70641 \\
       \hline
       B25 initial & 409 \\
       B25 uninformative age posterior cut & 275 \\
       B25 \teff\ cut & 253 \\
       B25 \logg\ cut & 217 \\
       B25 age cut & 189 \\
       \hline
       HU25-B25 cross-match & 141 \\
       HU25-B25 with full astrometry & 141 \\
       \hline
       \multicolumn{2}{c}{Total} \\
       \hline
       HU25-B20-B25 & 22184
    \end{tabular}
    \label{tab:stellar-cuts}
\end{table}

In order to calculate the maximum oscillation amplitude \zmax\ of a star, we also require its parallax, proper motion, and radial velocity, as well as the uncertainties for those parameters. We thus make one final cut to our sample, keeping only those stars with full astrometric solutions. For K2, this removes no stars from our HU25-B25 cross-match of 141 stars; for \textit{Kepler}, this takes our cross-matched HU25-B20 sample down from 50675 to 22043 (see Table \ref{tab:stellar-cuts}). In total, our final sample for this study comprises 22184 stars from \textit{Kepler} and K2. In the top panel of Figure \ref{fig:kiel-age-vs-height-hu25-b20-b25}, we show a Kiel diagram of these stars, colored by their isochrone ages, along with marginalized histograms of the \teff\ and \logg\ distributions. The median \teff\ in our sample is 5703 K (with a median absolute deviation, or MAD, of 215 K) and the median \logg\ is 4.356 dex (with a MAD of 0.089 dex). Our sample differs from that of Z23 primarily in our cuts on uninformative age posteriors, which removed most main sequence stars between 4000 and 5000 K. This is not surprising due to the difficulty in obtaining isochrone age measurements for cooler main sequence stars. This difference partially informed our addition to this study of the TRILEGAL stellar sample, which does contain cooler dwarf stars (see Section \ref{subsec:trilegal_stellar_sample}). 

\subsubsection{Age-\zmax\ relation}
\label{sec:age-zmax}

We then compute the vertical oscillation amplitude for each star in our curated sample using the software package \texttt{Gala} \citep{gala, adrian_price_whelan_2020_4159870}. Gala reconstructs galactic orbits from the instantaneous \textit{Gaia} kinematic information together with a model for the Milky Way galactic potential. We run \texttt{Gala} using a Galactic potential model of a spherical nucleus and bulge, a Miyamoto-Nagai disk, and a spherical Navarro-Frenk-White dark matter halo \citep{bovy_galpy_2015, navarro_structure_1996}. From these initial inputs and \textit{Gaia} DR3 parallax, proper motion, and radial velocity information, we integrate the orbits of each star in the sample for 250 Myr (approximately one Solar orbit around the Milky Way) with a timestep of 0.5 Myr before retrieving \zmax. 

\begin{figure}  
\includegraphics[width=.45\textwidth]{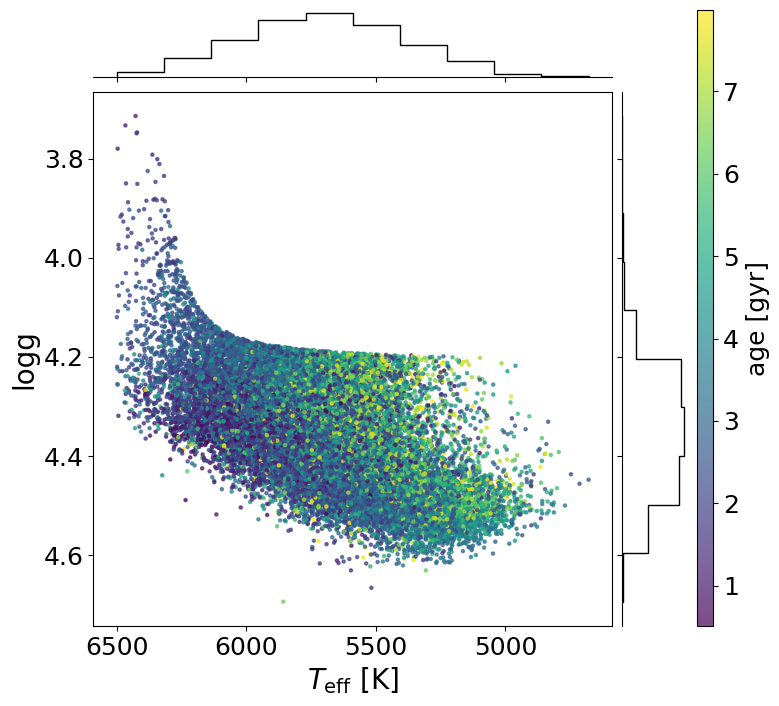}
\includegraphics[width=.45\textwidth]{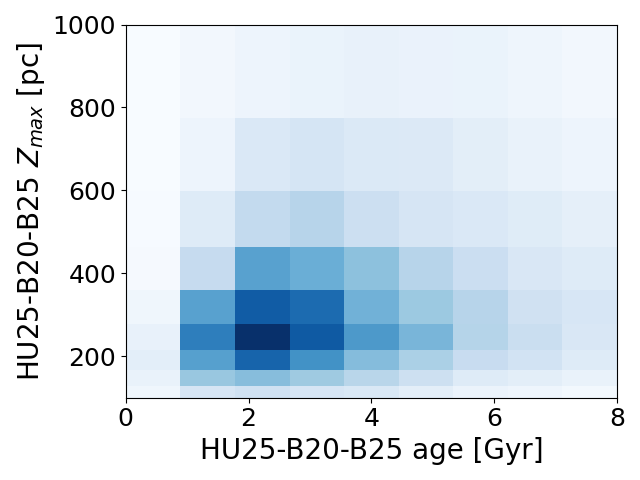}
\caption{\textbf{Top: }Kiel diagram of the combined HU25-B20-B25 sample, color-coded by age. Top and right marginalized histograms show the \teff\ and \logg\ distributions, respectively. The median \teff\ and \logg\ are 5703 K (MAD of 215 K) and 4.356 dex (MAD of 0.089 dex), respectively. \textbf{Bottom: }We observe a soft positive trend between isochrone age and $Z_{\textrm{max}}$, as calculated by the \texttt{Gala} \citep{gala, adrian_price_whelan_2020_4159870} software package.} 
\label{fig:kiel-age-vs-height-hu25-b20-b25}
\end{figure}

For a time-dependent planet formation history to produce a change in planet occurrence over \zmax, the stellar sample needs to exhibit a discernible trend between age and vertical oscillation amplitude. In the bottom panel of Figure \ref{fig:kiel-age-vs-height-hu25-b20-b25}, we show that using isochrone ages from our joint B20-B25-HU25 sample presents a soft positive trend consistent with previous investigations of \zmax\ for \textit{Kepler} stars \citep{miglio_age_2021, silva_aguirre_confirming_2018, casagrande_measuring_2016, sagear_zoomies_2024}. 


From this seed sample of 22184 \textit{Kepler} and K2 stars with ages and Galactic scale heights, we re-draw their ages and \zmax, in order to ensure that we properly account for the large stellar age uncertainties, along with their \teff, \logg, stellar radii, and stellar mass. We use the \textit{Gaia} DR3 isochrone-derived parameter uncertainties for each draw. We draw 30 stellar samples, around which we will synthesize 30 distinct planet populations (see Section \ref{sec:psps}). 

\subsubsection{TRILEGAL Kepler and K2 sample}
\label{subsec:trilegal_stellar_sample}

We next construct a synthetic joint \textit{Kepler}-K2 sample using TRILEGAL \citep{girardi_star_2005}, a population synthesis code for Milky Way stars. For our purposes, the TRILEGAL sample allows us to consider an optimistic scenario in which stellar ages are known precisely. To first order, a simulated TRILEGAL pointing of the \textit{Kepler} footprint furnishes a good match to measured ages from gyrochronology -- \citet{bouma_ages_2024} found that a two-step star formation rate (SFR) prescription, together with TRILEGAL's built-in kinematic heating model over time, is favored over a constant SFR and no kinematic heating, at least for a sample $<4$~Gyr. This motivates our choice of the same two-step SFR, in which the modeled thin disk produces a factor of 1.5 more stars with ages between 1--4 Gyr \citep{girardi_star_2005}. 

We compose the joint sample by synthesizing the \textit{Kepler} and K2 fields separately and then pieceing them together. To generate the TRILEGAL ``\textit{Kepler}-like" stellar sample, we query TRILEGAL\footnote{\url{http://stev.oapd.inaf.it/cgi-bin/trilegal}} with input pointing parameters of l=76.32 deg, b=13.5 deg, and field area of 10 $\textrm{deg}^2$. We employ the Chabrier lognormal initial mass function (IMF), binaries toggled on with a binary fraction of 0.5 (rather than the default of 0.3, based on the more recent results of \cite{offner_origin_2022}) and mass ratios of 0.7 to 1, default extinction parameters, a solar position of R=8122 pc and z=20.8 pc, a thin disk population generated using a two-step SFR (with otherwise default thin-disk parameters), a thick disk population generated using default parameters, and no halo or bulge population. TRILEGAL's prescription for kinematic heating includes an evolving scale height, with $h_z$ = $z_0$(1+$t$/$t_0$)\textsuperscript{$\alpha$}, where $z_0$=94.7 pc, $t_0$=5.55 Gyr, and $\alpha$ = 5/3). 

Since TRILEGAL parameters are generated on a grid, we perturb relevant stellar parameters (specifically, age and distance) with a spread equal to the bin size, following the procedure from \citet{bouma_ages_2024}. The bin sizes for log(age) and distance modulus are 0.02 and 0.05, respectively. This means that the average assigned age uncertainty for the TRILEGAL stellar sample is 1.0 Gyr; for comparison, the average upper and lower age uncertainties for the HU25-B20 \textit{Kepler} sample are 2.8 Gyr and 1.9 Gyr, respectively, while the average upper and lower age uncertainties for the HU25-B25 K2 sample are 5.6 Gyr and 3.4 Gyr. We build our initial TRILEGAL \textit{Kepler} sample by running the web tool ten times using the prescription described above, resulting in an initial sample size of 280,647 stars. This was subsequently cut to 132,951 non-binary stars, 101,635 stars with \teff\ between the maximum (6499 K) and minimum (4679 K) values in the combined \textit{Kepler}-K2 sample, 67,408 stars after the Equation \ref{eq:giant-cut} giant removal procedure was applied, 82177 stars between 4.0 and 4.7 in \logg, and 28,792 stars within the \textit{Kepler} magnitude range of HU25-B20-B25 (7.40-15.07). 

\begin{figure*}  
\includegraphics[width=.90\textwidth]{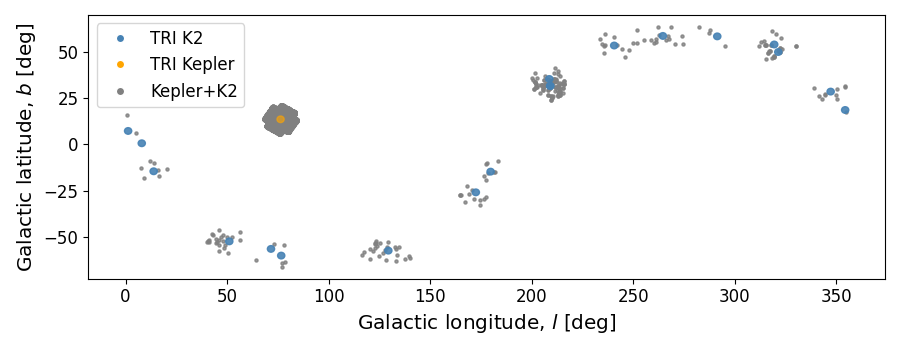}
\caption{Galactic latitudes and longitudes for our combined HU25-B20-B25 \textit{Kepler}$-$K2 sample (gray), the K2 TRILEGAL search field (blue), and the \textit{Kepler} TRILEGAL search field (orange). TRILEGAL search field dots are scaled to be 10 deg$^2$ (even if the fields themselves are not quite circles).} 
\label{fig:pointings}
\end{figure*}

Unlike \textit{Kepler}, the K2 mission was comprised of 19 separate campaigns (omitting Campaign 0) with different pointings. We make separate TRILEGAL queries for each of the 19 fields, two times each. Each query uses the same input parameters as the TRILEGAL \textit{Kepler} sample, except for the Galactic coordinates, which vary per campaign and are taken from the Data Release Notes on MAST\footnote{\url{https://archive.stsci.edu/missions-and-data/k2/documents/data-release-notes}}. In Figure \ref{fig:pointings}, we show the locations in Galactic latitude and longitude of the search fields for the \textit{Kepler} and K2 TRILEGAL samples, overlaid on the Galactic coordinates of our HU25-B20-B25 sample. The queries over the 19 campaigns returned 842,457 stars. 

\begin{figure*}  
\includegraphics[width=.90\textwidth]{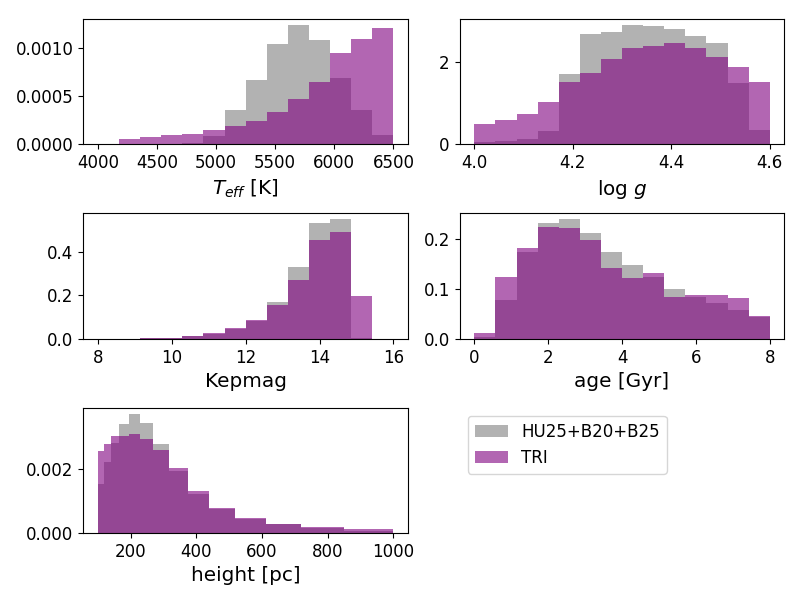}
\caption{Normalized histograms of key parameters for the HU25-B20-B25 \textit{Kepler}$-$K2 sample (gray) and the TRILEGAL \textit{Kepler}$-$K2 sample (purple). Note that the ``height" histogram refers to instantaneous height for TRILEGAL and \zmax\ for HU25-B20-B25. The medians and median absolute deviations (MAD) for each parameter are as follows: HU25-B20-B25 \teff\ = 5702$\pm$215 K; TRILEGAL \teff\ = 6053$\pm$272 K; HU25-B20-B25 \logg\ = 4.4$\pm$0.1 dex; TRILEGAL \logg\ = 4.4$\pm$0.1 dex; HU25-B20-B25 Kepmag = 14.0$\pm$0.5; TRILEGAL Kepmag = 14.1$\pm$0.6; HU25-B20-B25 age = 3.2$\pm$1.2 Gyr; TRILEGAL age = 3.2$\pm$1.4 Gyr; HU25-B20-B25 \zmax\ = 263$\pm$85 pc; TRILEGAL instantaneous height = 244$\pm$107 pc.} 
\label{fig:tri-vs-sample}
\end{figure*}

From this initial set, we cut the sample to 496,124 non-binary stars, 131,346 stars with \teff\ between the maximum and minimum values of the HU25-B20-B25 sample, 61,736 stars after applying the giant cut with Equation \ref{eq:giant-cut}, and 20,317 stars with \textit{Kepler} magnitude within the HU25-B20-B25 sample's corresponding range. In Figure \ref{fig:tri-vs-sample}, we compare the \teff, \logg, \textit{Kepler} magnitude, age, and instantaneous height (or \zmax, where applicable) distributions of our TRILEGAL and HU25-B20-B25 samples. Notably, HU25-B20-B25 peaks approximately 350 K cooler than TRILEGAL, and we attribute this to the uninformative age posterior cuts described above. The TRILEGAL \logg\ distribution is also wider, with more puffy stars that are likely sub-giants and early post-main sequence giants, as well as more high-\logg\ stars in the main sequence. In age (and Galactic height), HU25-B20-B25 also peaks slightly later (and higher). In total, across the \textit{Kepler} and K2 field queries, our final TRILEGAL sample contains 49,109 stars.


\begin{figure}  
\includegraphics[width=.45\textwidth]{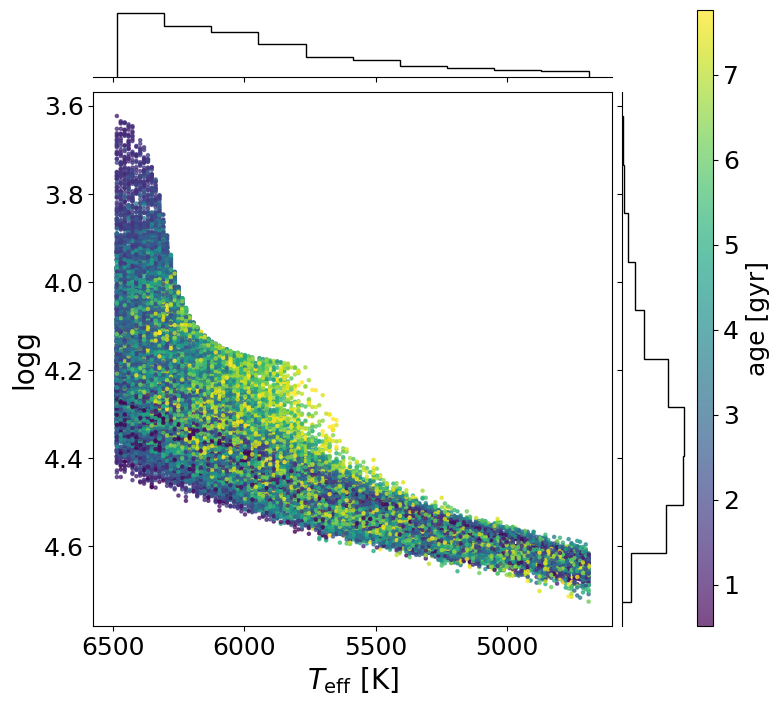}
\includegraphics[width=.45\textwidth]{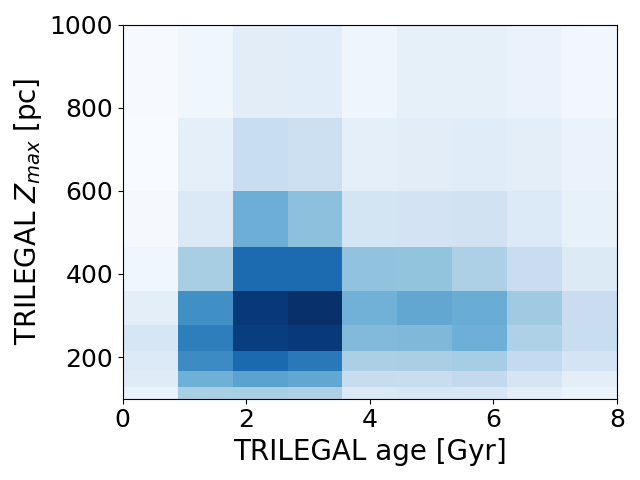}
\caption{\textbf{Top: }Kiel diagram of the TRILEGAL \textit{Kepler}-K2 sample, color-coded by age. Top and right marginalized histograms show the \teff\ and \logg\ distributions, respectively. \textbf{Bottom: }Age-\zmax\ relation for the TRILEGAL \textit{Kepler}-K2 sample. Note that height for the TRILEGAL is the instantaneous height, rather than \zmax.} 
\label{fig:kiel-age-height-tri-kepler}
\end{figure}

A Kiel diagram for the TRILEGAL sample is shown in top panel of Figure \ref{fig:kiel-age-height-tri-kepler}. There are two main differences between this sample and our HU25-B20-B25 based on real \textit{Kepler} and K2 stars. First, the TRILEGAL sample extends well into the K dwarf regime. Second, it omits likely sub-giants in the 5000 < \teff\ < 5500 K and 4.2 < \logg\ 4.4 dex range.

Unlike the HU25-B20-B25 sample, the TRILEGAL stellar sample does not have astrometric parameters with which to calculate \zmax. Instead, we are given the distance modulus, which, together with the inclination of the \textit{Kepler} field, can be used to calculate each star's instantaneous ``height" from the midplane. In bottom panel of Figure \ref{fig:kiel-age-height-tri-kepler}, we show that the TRILEGAL stellar sample exhibits a positive relation between age and this height, similar to our ``real" sample (bottom panel of Figure \ref{fig:kiel-age-vs-height-hu25-b20-b25}). By definition, one should expect the TRILEGAL instantaneous heights to be lower than the B20 \zmax\ values; this is borne out in our samples, although only slightly, with the TRILEGAL median height being 244 pc (MAD of 107 pc) and the HU25-B20-B25 median \zmax\ being 263 pc (MAD of 85 pc). 


\subsection{Time-dependent occurrence model}
\label{sec:models}

We employ two families of functions to model the planet occurrence rate history of the Milky Way. In both prescriptions, we assume that the Milky Way is 13.7 Gyr old \citep{xiang_time-resolved_2022, gallart_uncovering_2019}, and that planet formation has proceeded during that time in multiple phases. An acceptable model must also reproduce the present-day planet-host fraction of $\sim$30\% \citep{zhu_about_2018} of stars, though the path to arriving at that fraction can vary. Given that we are attempting to mimic a late-time rise in planet occurrence consistent with a higher midplane fraction, planet occurrence increases monotonically as the galaxy ages in all models.

The first family of models we explore are step functions with three parameters: 

\begin{itemize}
  \item fraction of stars hosting a planetary system before a planet formation-boosting threshold, \fone
  \item fraction of stars hosting a planetary system after the planet formation-boosting threshold, \ftwo
  \item lookback time at which \fone\ increases to \ftwo, \threshold.
\end{itemize}

\begin{figure}
\includegraphics[width=.45\textwidth]{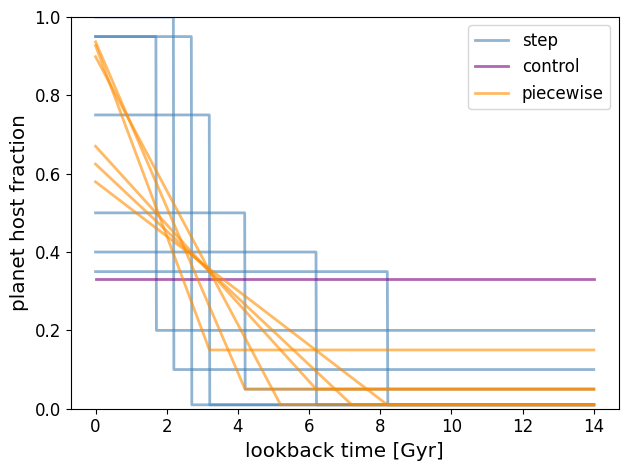} 
\caption{Galactic sculpting models, characterized by step and piecewise functions with a lower planet-host fraction after some event in the Milky Way's past led to an increase in planet hosts among Sun-like stars. We also include a control model that holds the planet host fraction constant over time.} 
\label{fig:models}
\end{figure}

A model of \{\fone=0.20, \ftwo=0.95, and \threshold=1.7\}, for example, means 20\% of stars hosted planets up until $\sim$1.7~Gyr ago (that is, in the first $\sim$12~Gyr of the Milky Way's lifetime). After that point, 95\% of stars host a planetary system. In Figure \ref{fig:models}, we show seven different step-function models, with \threshold\ ranging from 1.7 to 8.2~Gyr ago and \fone\ and \ftwo\ chosen to yield an overall present-day planet host fraction of 30-35\% \citep{zhu_about_2018}, as well as a control (flat) model set constant at \f=33\%. The step function models are as follows:
\begin{enumerate}
    \item \fone=20\%, \ftwo=95\%, \threshold=1.7 Gyr ago
    \item \fone=10\%, \ftwo=100\%, \threshold=2.2 Gyr ago
    \item \fone=1\%, \ftwo=95\%, \threshold=2.7 Gyr ago
    \item \fone=1\%, \ftwo=75\%, \threshold=3.2 Gyr ago
    \item \fone=5\%, \ftwo=50\%, \threshold=4.2 Gyr ago
    \item \fone=1\%, \ftwo=40\%, \threshold=6.2 Gyr ago
    \item \fone=1\%, \ftwo=35\%, \threshold=8.2 Gyr ago
\end{enumerate}



In the second family of models for planet occurrence with time, rather than an instantaneous jump between two constants, we employ a piecewise function. Planet occurrence begins with an initially constant planet-host fraction, \fone, followed by a positively sloped linear function of \f\ with cosmic time (or, equivalently, a negatively sloped function of planet occurrence with lookback time or stellar age), up to a constant, present-day planet-host fraction, \ftwo. We choose the piecewise functions such that they increase until the present-day in order to construct a model archetype that is sufficiently different from the step function. As with the step function models, these piecewise models are chosen such that the total planet host fraction for stars of all ages comes to 30-35\%. The piecewise models employed in this study are as follows and are illustrated in orange in Figure \ref{fig:models}:
\begin{enumerate}
    \item \fone=15\%, \ftwo=100\%, \threshold=3.2 Gyr ago
    \item \fone=5\%, \ftwo=100\%, \threshold=4.2 Gyr ago
    \item \fone=1\%, \ftwo=95\%, \threshold=5.2 Gyr ago
    \item \fone=5\%, \ftwo=70\%, \threshold=6.2 Gyr ago
    \item \fone=1\%, \ftwo=65\%, \threshold=7.2 Gyr ago
    \item \fone=1\%, \ftwo=60\%, \threshold=8.2 Gyr ago,
\end{enumerate}

For both model archetypes, model selection began with a wide range of \threshold\ and iterated to the set presented in this study. Each \threshold\ is degenerate in that there is a set of \fone\ and \ftwo\ that lead to a present-day \f\ of 30-35\%. As we will later see in Section \ref{sec:results}, however, so much of the model space is ruled out due to a too-shallow occurrence-\zmax\ trend. Therefore, we can rule out entire \threshold\ by testing the two solutions per \threshold\ that generate the theoretically steepest trend -- one with \fone=1\% and one with \ftwo=100\% -- and finding that both are still too shallow. 


\subsection{Planetary System Population Synthesis}
\label{sec:psps}

We now have in hand a stellar sample, spanning a variety of ages, with some established fraction hosting planets according to the prescription described in Section \ref{sec:models}. We have already established that acceptable models must reproduce the present-day planet host fraction of $\sim$30-35\%. In order to maintain consistency with observed transit multiplicity as well, we adopt a heuristic for planetary system architecture. There is theoretical support for planetary systems exhibiting warmer dynamical temperatures with stellar age \citep{pu_spacing_2015, volk_consolidating_2015}, but \citet{lam_ages_2024} found no distinction over Gyr timescales between dynamically hot and cold systems from the same B20 sample. They concluded rather that $18^{+15}_{-10}$\% of planet-hosting \textit{Kepler} FGK dwarfs have a ``system of tightly-packed inner planets" (STIPs) characterized by low angular momentum deficit \citep{he_architectures_2020}, but with no evidence that the fraction changes with time. We designate systems as either dynamically ``hot" or ``cold" via their system properties such as mutual inclination, orbital eccentricity, and planet multiplicity \citep{tremaine_statistical_2015, he_architectures_2020, laskar_amd-stability_2017}. We hold the relative hot/cold fraction to be constant, varying only the fraction of stars that host a planetary system. In other words, stellar age determines only the fraction of planet hosts in the sample and not the number of planets per system, nor their transit geometries.

We determine the fraction of dynamically ``cool" systems in each stellar sample by drawing from a normal distribution, $\sim$$N$(0.18, 0.1), which approximates the corresponding posterior distribution from \citet{lam_ages_2024}. Planet-hosting systems designated dynamically cool are randomly assigned either 5 or 6 planets, with inclinations from the midplane drawn from $N$($\mu$=0\degree, $\sigma$=2\degree) \citep{ballard_kepler_2016, dawson_correlations_2016, zhu_about_2018, he_architectures_2020, zawadzki_migration_2022, lam_ages_2024}. Systems that are dynamically hot are randomly assigned either 1 or 2 planets, with inclinations from the midplane drawn from $N$($\mu$=0\degree, $\sigma$=8\degree). Planet eccentricities are drawn from a Rayleigh distribution with a peak at 0.24 for singles and a peak at 0.06 for multis \citep{eylen_orbital_2019}. All planet hosts are assigned a midplane orientation randomly drawn from $U$(-$\pi$/2, $\pi$/2). All planets are assigned a longitude of periastron that is also randomly drawn from $U$(-$\pi$/2, $\pi$/2). 

Planets are assigned equal probability of being either a Super-Earth (1.2 $R_{\oplus}$ < $R_{p}$ < 2 $R_{\oplus}$) or Sub-Neptune (2 $R_{\oplus}$ < $R_{p}$ < 4 $R_{\oplus}$). As in Z23, the radii are assigned following a power law:
\begin{equation}
    q(R_p) = R_p^\alpha,
    \label{eq:p_intact3}
\end{equation}
where $R_{p}$ is the planet radius and $\alpha$ is drawn once per stellar population from $N$($\mu$=-1, $\sigma$=0.2) if the planet is a Super-Earth (1.2 $R_{\oplus}$ < $R_{p}$ < 2 $R_{\oplus}$) and from $N$($\mu$=-1.5, $\sigma$=0.1) if the planet is a Sub-Neptune (2 $R_{\oplus}$ < $R_{p}$ < 4 $R_{\oplus}$). We then draw planet periods on either side of the radius valley as parameterized by \cite{vaneylen_asteroseismic_2018}, which found a slope $\gamma$ of -0.09$^{+0.02}_{-0.04}$, upper envelope y-intercept $a_{upp}$ of 0.44$^{+0.04}_{-0.03}$, and lower envelope y-intercept $a_{low}$ of 0.29$^{+0.04}_{-0.03}$. Planets are given a 5\% chance of falling within the valley. For each of our 30 stellar population realizations, we draw $\gamma$, $a_{upp}$, and $a_{low}$ from asymmetric distributions following the approximate method in \cite{jontof-hutter_following_2021} and \cite{barlow_asymmetric_2004}. Masses are ascribed using the \texttt{forecaster} radius-mass relation \citep{chen_probabilistic_2016}. We calculate mutual Hill stability and redraw if the system is Hill unstable \citep{chambers_stability_1996, smith_orbital_2009, fabrycky_architecture_2014}. Finally, to place our planet sample on even footing with Z23, we keep only planets with orbital period less than 40 days, to compare against Z23. 

\subsection{Detection and Completeness}
\label{sec:completeness}
After generating planetary systems around the stellar sample, we ``observe" the sample to determine the transit yield. We determine which of the planets geometrically transit, based on the orbital parameters drawn in Section \ref{sec:psps}. To do so, we use Equation 7 in \cite{winn_exoplanet_2010}:
\begin{equation}
    b = \frac{a*\cos{i}}{R_*} \frac{1-e^2}{1+e*\sin{\omega}},
\label{eq:impact_parameter}
\end{equation}
where $a$ is the semi-major axis, $i$ is the mutual inclination of the planet's orbit from its system's midplane, $e$ is the planet's orbital eccentricity, and $\omega$ is the longitude of periastron. 

Next, we use two different prescriptions of sensitivity, to respect the difference in the sensitivity functions of \textit{Kepler} and K2. For \textit{Kepler}, the signal to noise ratio (SNR) is calculated following Equation 4 in \cite{christiansen_derivation_2012}:
\begin{equation}
    SNR = \sqrt{\frac{t_{obs}*f_0}{P}} \frac{(R_p/R_*)^2}{CDPP_{eff}}, 
\label{eq:snr}
\end{equation}
where 
\begin{equation}
    CDPP_{eff} = \sqrt{\frac{t_{CDPP}}{t_{dur}}}{CDPP_N}.
\label{eq:cdpp_eff}
\end{equation}

We then apply the SNR versus detection probability ramp from \citet{fressin_false_2013} to establish a detection likelihood (see also Section 3.3 of \citet{lam_ages_2024} for an identical application). Each geometrically transiting planet is assigned a detection likelihood, upon which a binary detection or non-detection is drawn. 

For K2, we follow the prescription the recovery fraction, $f(MES)$, from Equations 5 and 6 in \cite{zink_scaling_2021}:
\begin{equation}
    MES = C\frac{depth}{CDPP_{t_{dur}}} \sqrt{N_{tr}}, 
\label{eq:k2-mes}
\end{equation}
and
\begin{equation}
    f(MES) = \frac{a}{1+e^{-k(MES-1l}},
\label{eq:k2-recovery-fraction}
\end{equation}
where $CDPP_{t_{dur}}$ = $CDPP_{6 hr}$ (obtained from Table 2 of \cite{zink_scaling_2021}), $N_{tr}$ is calculated using the baseline of the particular K2 campaign and the planet's period, depth is calculated using the stellar and planet radii, $C$ is a global correction factor of 0.9488, and a, k, and l are best-fit parameters for the logistic function describing K2's recovery fraction. These logistic function parameters are provided in Table 3 of \cite{zink_scaling_2021} for Campaigns 1-8 and 10-18, while for Campaigns 9 and 19, we use the summary values marginalized over AFGK dwarfs ($a$=0.6095, $k$=0.6088, $l$=10.8986). As with the detection likelihood in our \textit{Kepler} sample, we draw a detection or non-detection for each geometrically transiting K2 planet based on its recovery fraction.

In this study, we use the inverse detection efficiency method (IDEM), in which the detection efficiency is independently computed and then applied to a detected yield in order to estimate a ``true" or ``adjusted" yield. The detection efficiency is computed along 9 bins each of planet period (log space from 1-40 days) and radius (linear space from 1-4 $R_{\oplus}$), and is a product of the correspondingly-binned detection sensitivity and geometric transit probability maps. The geometric transit probability map is constructed once, since it is just a function of the planet semi-major axis and stellar and planet radii, which we assume to vary negligibly over \zmax. It is evaluated at each period and radius bin as $(R_* + R_p)/a$, where $a$ is the planet semi-major axis, and $R_*$ is assumed to be 1 $R_{\odot}$. The period is converted to $a$ using Kepler's Third Law assuming 1 $M_{\odot}$.

\begin{figure*}
  {\begin{tabular}{@{}cc@{}}
    \includegraphics[width=0.49\textwidth, height=9.5cm]{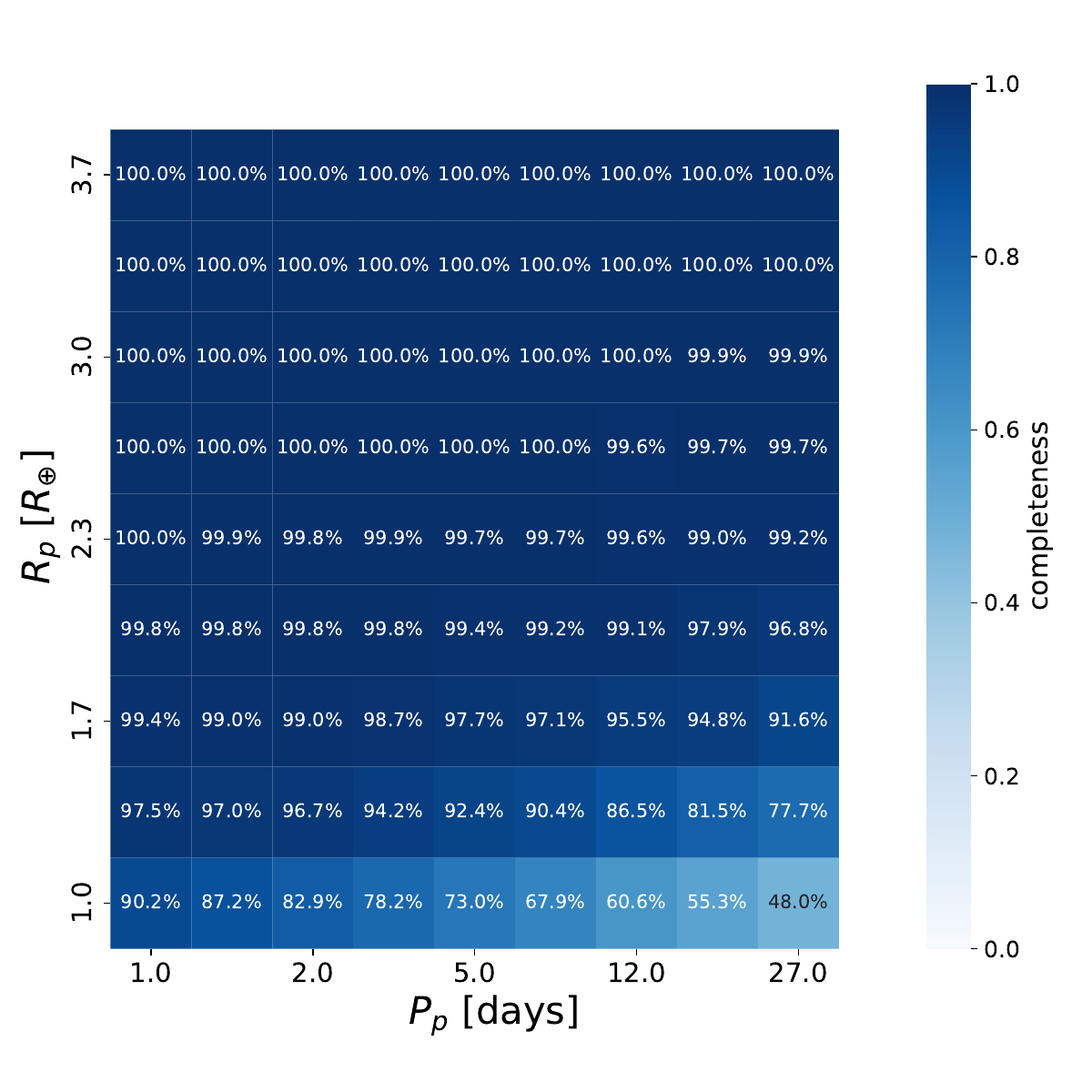}&
    \includegraphics[width=0.49\textwidth, height=9.5cm]{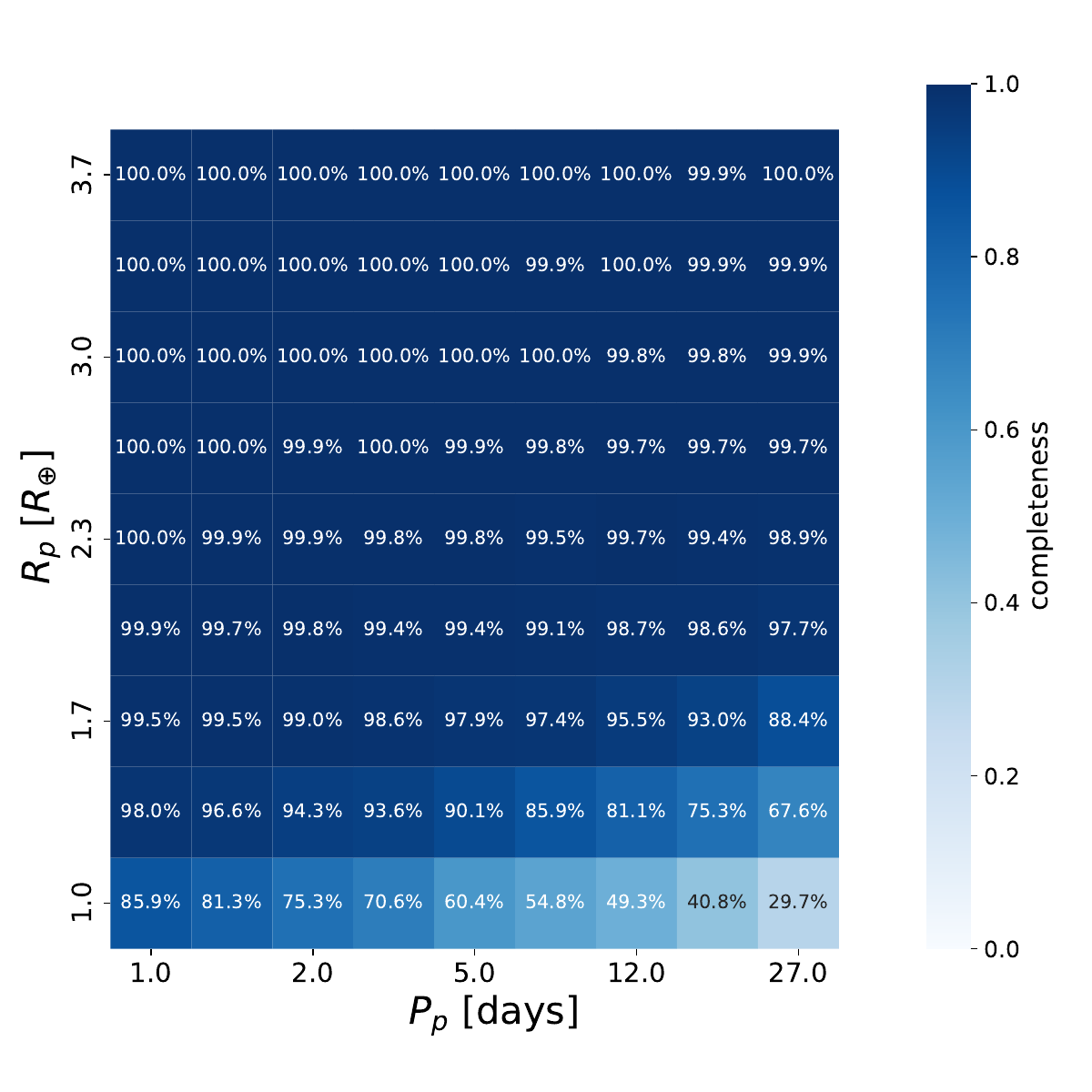}\\
    \includegraphics[width=0.49\textwidth, height=9.5cm]{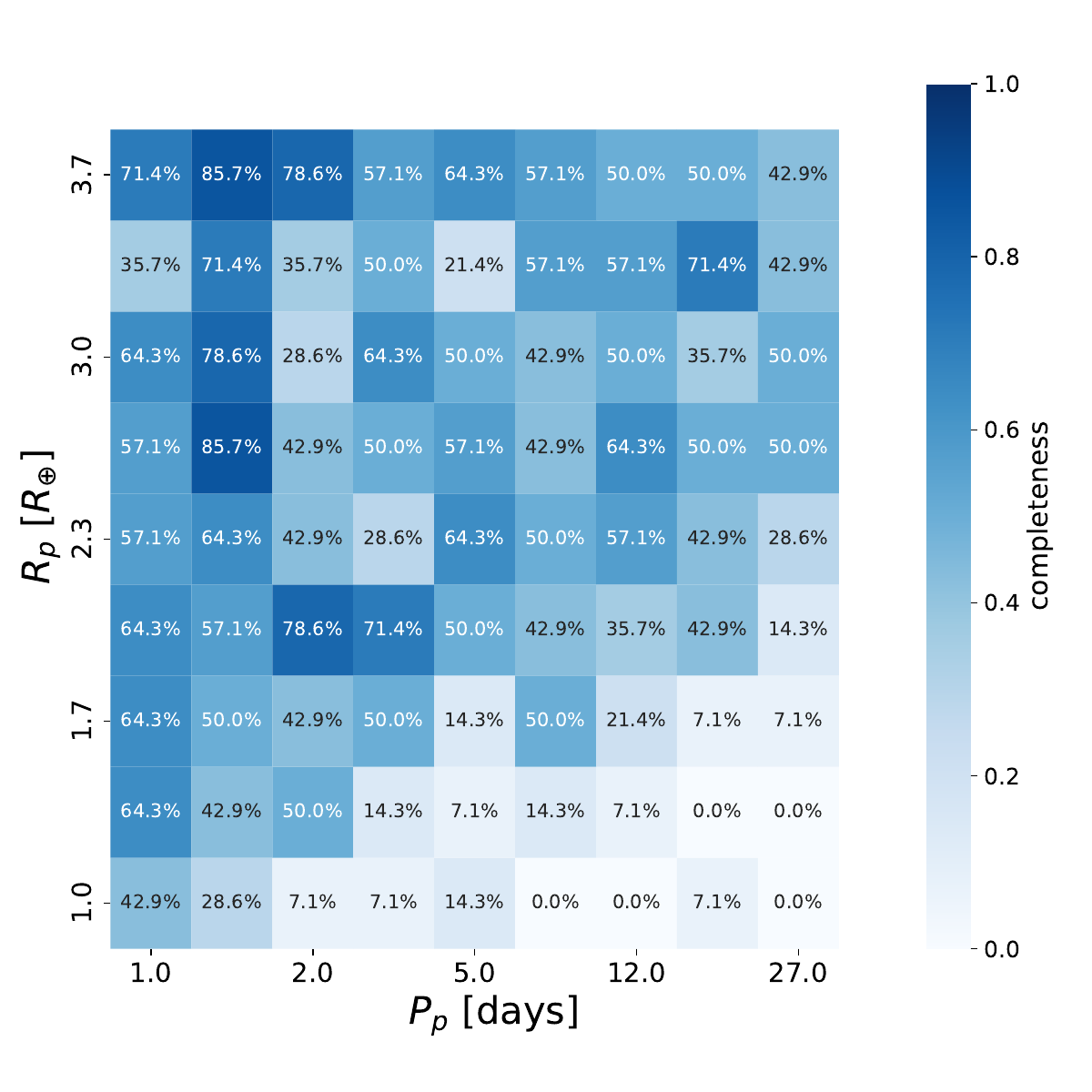}&
    \includegraphics[width=0.49\textwidth, height=9.5cm]{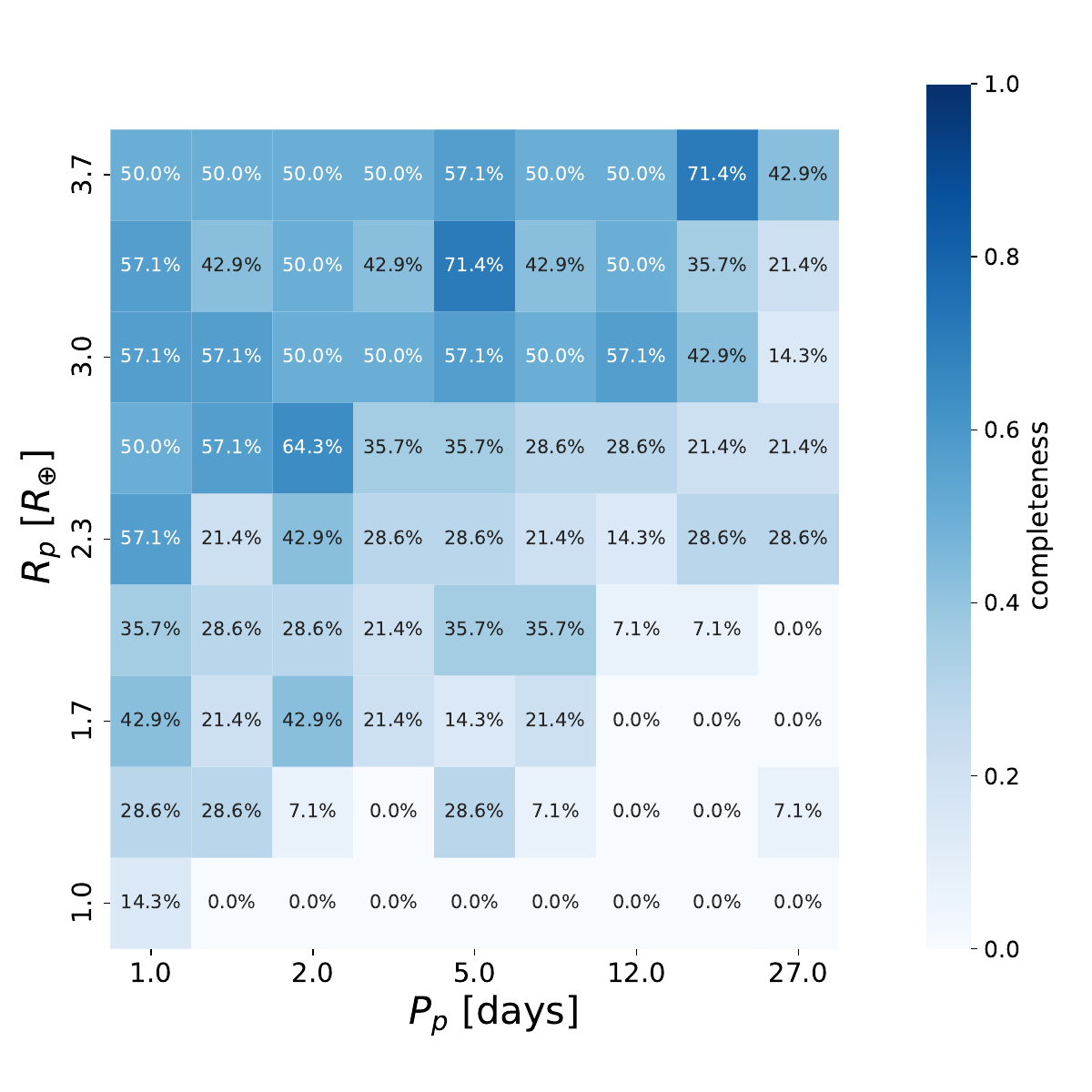}
  \end{tabular}}
    \caption{Detection sensitivity maps were constructed by injecting a geometrically transiting planet at each period and radius bin for 1000 random \textit{Kepler} stars or the entirety of the K2 stars in each \zmax\ bin. Here we show four representative sensitivity maps: lowest-\zmax\ bin \textit{Kepler} stars (from 100-158 pc) in the top left, highest-\zmax\ bin \textit{Kepler} stars (from 631-1000 pc) in the top right, lowest-\zmax\ bin K2 stars in the bottom left, and highest\zmax\ bin K2 stars in the bottom right.}
  \label{fig:sensitivity}
\end{figure*}

Meanwhile, a detection sensitivity map is constructed for each of the five \zmax\ bins (in log space from 100 to 1000 pc). We take 1000 random stars from each height bin of the \textit{Kepler} sample (and all of the stars from each height bin in the K2 sample, since there are so few of such stars), injecting a planet with an impact parameter of zero at each period and radius bin. We next determine which transiting planets exceed the detection threshold by applying the \textit{Kepler} and K2 sensitivity functions described above. The detection sensitivity in each period and radius bin is the number of detected planets out of 1000 (or however many K2 planets there are in that particular height bin). As fiducial examples, we show the \textit{Kepler} and K2 detection sensitivity maps for the lowest and highest \zmax\ bins in Figure \ref{fig:sensitivity}. The K2 sensitivity maps are dominated by shot noise due to the much smaller sample, particular in bins where both sensitivity and the planet count is low; this is a known problem in real occurrence calculations and an inherent instability of the inverse detection efficiency method \citep{hsu_improving_2018}. We check and verify that using these versus the sensitivity maps from \cite{thompson_planetary_2018} and \cite{zink_scaling_2021} for \textit{Kepler} and K2, respectively, make no appreciable difference to the adjusted planet yields. 

\section{Results}
\label{sec:results}

\subsection{Results Using Joint Kepler-K2 Sample}
\label{sec:results-joint}

The uncertainty at each \zmax\ bin reflects the apparent change in planet yield across 30 draws from the completeness function. This uncertainty per \zmax\ bin depends upon the model occurrence prescription, as it ought: depending on when in time planetary systems are born, they will move between \zmax\ bins according to kinematic heating. In draws with smaller numbers of planets in that \zmax\ bin, the resulting occurrence estimate is noisier.

We use the NUTS sampler in the \texttt{numpyro} framework \citep{phan_composable_2019} to fit a simplified version of Equation 4 in Z23 to the recovered yield. The model is constructed as follows: 
\begin{equation}
    y = 100 * \eta * Z_{\textrm{max}}^{\tau} * \frac{1}{C * dlnZ_{max}},
    \label{eq:power-law}
\end{equation}

with normalizing parameter
\begin{equation}
    C = \frac{1}{\tau+1}1000^{\tau+1} - \frac{1}{\tau+1}100^{\tau+1},
    \label{eq:power-law-constant}
\end{equation}

and normalizing bin-size dln\zmax\ = 0.0011. Our support for \zmax\ spans 100 to 1000 pc, while $y$ represents the planet occurrence rate per 100 stars. The free parameters, $\tau$ and $\eta$, represent the slope of the planet occurrence-scale height relation and occurrence rate normalization, respectively. $\tau$ is drawn from a uniform prior of $U(-1, 1)$ (that is, we allow for the model to fit cases where planet occurrence \textit{increases} with \zmax), while $\eta$ is drawn from a uniform prior of $U(0.01, 1.0)$. We run the MCMC for 1000 warm-up steps, 10000 samples, and 8 chains (making sure acceptance rates were at least 90\%), sampling from a normal distribution around $y$ with error equal to the per-height-bin standard deviation of the completeness-recovered planet occurrence yield. 

\begin{figure*}
  {\begin{tabular}{@{}cc@{}}
    \includegraphics[width=0.47\textwidth, height=5.25cm]{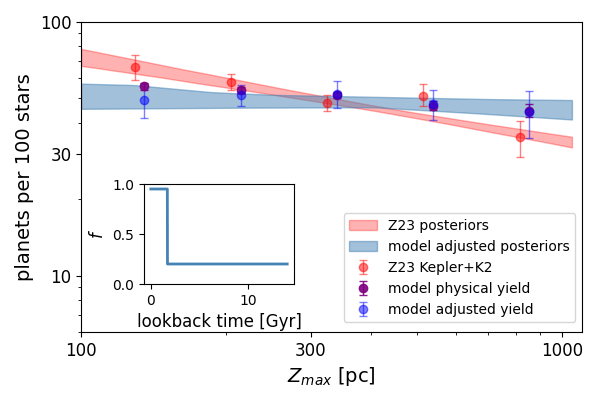}&
    \includegraphics[width=0.47\textwidth, height=5.25cm]{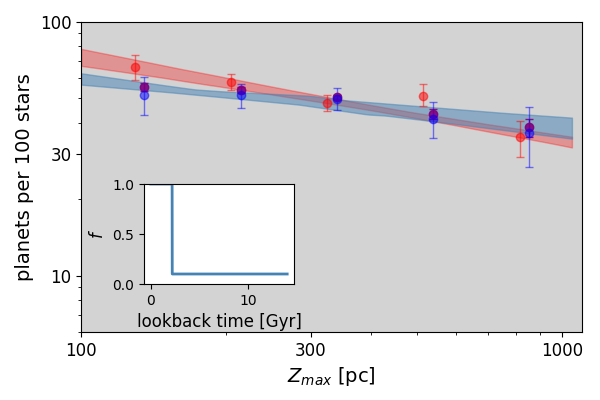}\\
    \includegraphics[width=0.47\textwidth, height=5.25cm]{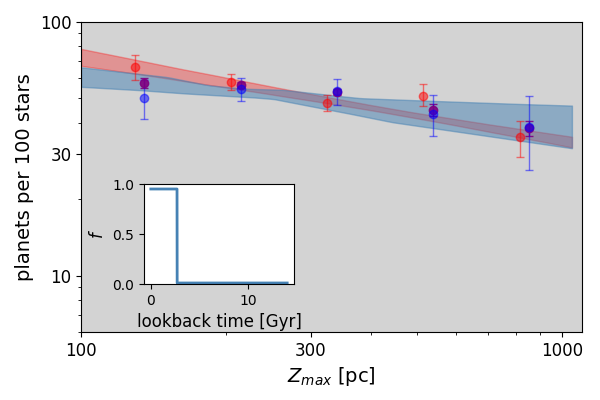}&
    \includegraphics[width=0.47\textwidth, height=5.25cm]{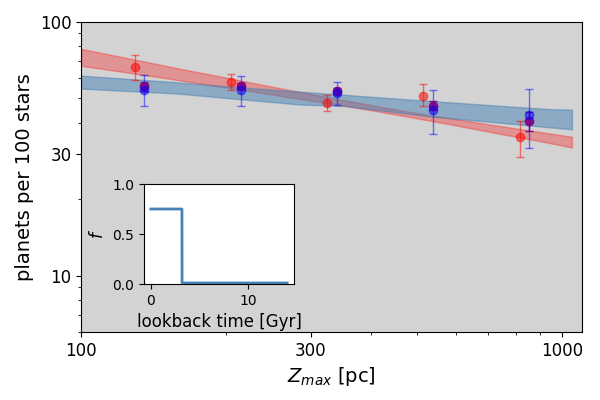}\\
    \includegraphics[width=0.47\textwidth, height=5.25cm]{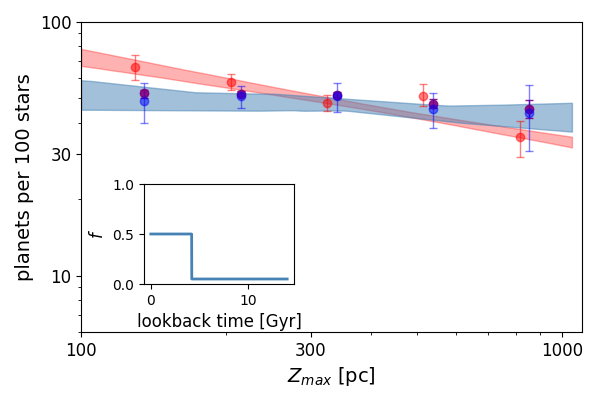}&
    \includegraphics[width=0.47\textwidth, height=5.25cm]{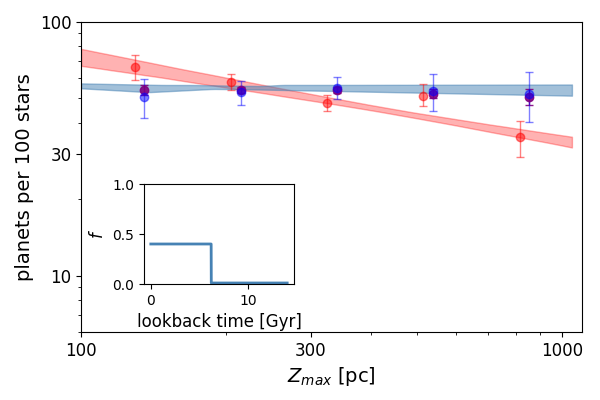}\\
    \includegraphics[width=0.47\textwidth, height=5.25cm]{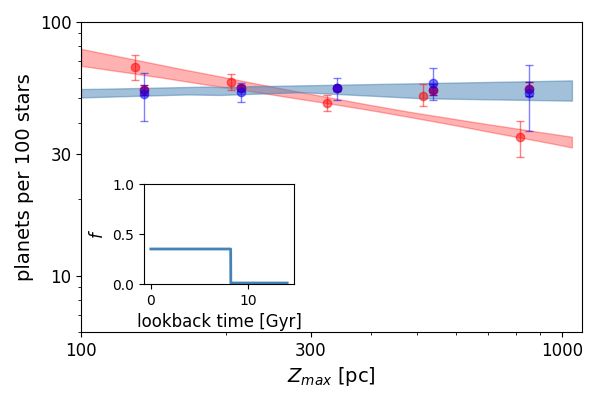}&
    \includegraphics[width=0.47\textwidth, height=5.25cm]{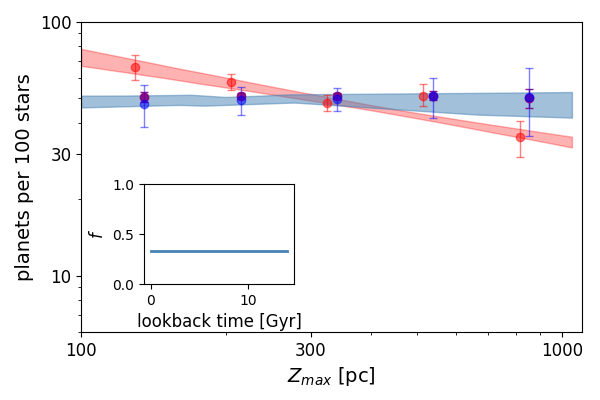}
  \end{tabular}}
    \caption{Planet occurrence versus \zmax\ for the step function planet occurrence models, shown in insets. We consider only planets with period 1 < P\textsubscript{p} < 40 days and radius 1.2 $R_{\oplus}$ < R\textsubscript{p} < 4 $R_{\oplus}$. All models are constrained to produce a present-day planet host fraction of approximately 0.3. Red points and envelope correspond to the \textit{Kepler} small planet occurrence reported by Z23 and the best fit, respectively. Blue points and envelope correspond to the model's completeness-adjusted planet occurrence rate. Envelopes correspond to 16th and 84th percentiles. Purple points correspond to the ``true" physical planet occurrence rate from the model. Markers are offset for clarity. Favored models are shaded in gray. \textbf{Top left:} \fone=20\%, \ftwo=95\%, \threshold=1.7 Gyr. \textbf{Top right:} \fone=10\%, \ftwo=100\%, \threshold=2.2 Gyr. \textbf{Second row left:} \fone=1\%, \ftwo=95\%, \threshold=2.7 Gyr. \textbf{Second row right:} \fone=1\%, \ftwo=75\%, threshold=3.2 Gyr. \textbf{Third row left:} \fone=5\%, \ftwo=50\%, \threshold=4.2 Gyr. \textbf{Third row right:} \fone=1\%, \ftwo=40\%, \threshold=6.2 Gyr. \textbf{Fourth row left:} \fone=1\%, \ftwo=35\%, \threshold=8.2 Gyr. \textbf{Fourth row right:} Planet host fraction constant at 33\%.}
  \label{fig:model-yields-new}
\end{figure*}

In Figure \ref{fig:model-yields-new}, we show the resulting occurrence-\zmax\ trend for the step function models illustrated in blue in Figure \ref{fig:models}, overlaid against the trend found by Z23 (in red). Z23 analyzed Super-Earths and Sub-Neptunes separately; in this work, we combine these into one designation of ``rocky" planets from 1 to 4 \rearth, eschewing our ability to comment on the age dependence of the radius valley, in favor of greater statistical power for our toy models. The physical (``true", in  dark purple) and recovered (blue) planet occurrences from each model are binned in the same manner as in Z23 -- five evenly log-spaced bins from 100 to 1000 pc. We show the 16th and 84th percentile envelope fits to our recovered yields with the blue envelopes in Figure \ref{fig:model-yields-new}, and we report the best-fit planet occurrence trend slope ($\tau$) and occurrence normalization ($\eta$) for each model in Table \ref{tab1}. For comparison, Z23 found $\tau$ = -0.30$\pm$0.06 for Super-Earths, and $\tau$ = -0.37$\pm$0.07 for Sub-Neptunes. After combining these separate observed samples \footnote{Courtesy of J. Zink.} and adding their uncertainties in quadrature, we re-fit and find their $\tau$ to be -0.28$\pm$0.08 and $\eta$ to be 0.46$\pm$0.02 for small, close-in planets. The 16th and 84th percentile envelopes for this fit are also plotted in red in Figure \ref{fig:model-yields-new}. 

\begin{table}
\centering
  \caption{Planet occurrence trend slopes for different models }
    \begin{tabular}{ c|c|c }
    Step function model & $\tau$ & $\eta$\\ 
     \hline
     \{\fone=20\%, \ftwo=95\%, \threshold=1.7 Gyr\} & -0.05 $\pm{0.10}$ & 0.48 $\pm{0.04}$\\ 
     \{\fone=10\%, \ftwo=100\%, \threshold=2.2 Gyr\} & -0.17 $\pm{0.12}$ & 0.44 $\pm{0.04}$\\  
    \{\fone=1\%, \ftwo=95\%, \threshold=2.7 Gyr\} & -0.13 $\pm{0.13}$ & 0.47 $\pm{0.04}$\\  
    \{\fone=1\%, \ftwo=75\%, \threshold=3.2 Gyr\} & -0.10 $\pm{0.12}$ & 0.47 $\pm{0.04}$\\  
     \{\fone=5\%, \ftwo=50\%, \threshold=4.2 Gyr\} & -0.07 $\pm{0.12}$ & 0.47 $\pm{0.04}$\\  
      \{\fone=1\%, \ftwo=40\%, \threshold=6.2 Gyr\} & 0.02 $\pm{0.12}$ & 0.53 $\pm{0.04}$\\
      \{\fone=1\%, \ftwo=35\%, \threshold=8.2 Gyr\} & 0.05 $\pm{0.13}$ & 0.55 $\pm{0.05}$\\
      \{\fone=33\%, \ftwo=33\%, \threshold=N/A\} & 0.05 $\pm{0.14}$ & 0.50 $\pm{0.05}$\\
     \hline
     Piecewise function model & $\tau$ & $\eta$\\
     \hline
     \{\fone=15\%, \ftwo=100\%, \threshold=3.2 Gyr\} & -0.06 $\pm{0.10}$ & 0.48 $\pm{0.03}$\\
     \{\fone=5\%, \ftwo=100\%, \threshold=4.2 Gyr\} & -0.08 $\pm{0.14}$ & 0.45 $\pm{0.04}$\\
     \{\fone=1\%, \ftwo=95\%, \threshold=5.2 Gyr\} & -0.10 $\pm{0.11}$ & 0.49 $\pm{0.04}$\\
     \{\fone=5\%, \ftwo=70\%, \threshold=6.2 Gyr\} & -0.05 $\pm{0.12}$ & 0.48 $\pm{0.04}$\\
     \{\fone=1\%, \ftwo=65\%, \threshold=7.2 Gyr\} & -0.02 $\pm{0.11}$ & 0.49 $\pm{0.04}$\\
     \{\fone=1\%, \ftwo=60\%, \threshold=8.2 Gyr\} & -0.01 $\pm{0.13}$ & 0.50 $\pm{0.05}$\\ \\
    \end{tabular}  
  \label{tab1}
\end{table}

Our control (``no change whatsoever") model, in which planet occurrence is stalled at 33\% for the age of the Milky Way, results in a constant planet occurrence rate at all galactic heights and is unsurprisingly inconsistent with the Z23 results. However, we find moreover that for the majority of our step function models, the occurrence-\zmax\ trend is similarly insufficiently steep. In general, \fone\ controls the high-\zmax\ end of the trend and \ftwo\ controls the low-\zmax\ end. Since the present-day total planet host fraction is fixed, it therefore follows that the limit at \fone=1\% and \ftwo=100\% sets the maximum slope of the occurrence-\zmax\ trend for any given \threshold. Using this conceit, it is possible to systematically rule out \threshold\ because these limiting cases fail to produce a sufficiently steep trend. The models that are not ruled out have step thresholds \threshold=2.2-3.2 Gyr ago. The models with \threshold=2.7 and \threshold=3.2 Gyr are contrived to push the possible window of step function model thresholds as wide as possible, with the former reaching the theoretical maximum \ftwo\ and the latter reaching the theoretical minimum \fone. In so doing, we can rule out models in which the Galactic small planet host fraction around FGK dwarfs jumped (in a relatively short time) any earlier than 2.2 Gyr ago or any later than 3.2 Gyr ago. 

In general, models that fail to match Z23 do not produce the necessary overabundance of planets with maximum oscillation amplitudes close to the Galactic midplane and relative underabundance of planets that travel to higher vertical excursions from the Galactic midplane, instead producing a flatter slope across the relevant \zmax\ range. This issue persists among the piecewise models, which are parameterized as a flat planet host fraction \fone\ until some time \threshold\ at which the fraction of planet hosts gradually increases to a present-day level of \ftwo. For this class of models, we rule out those with \threshold=6.2 Gyr ago or earlier, as well as \threshold=3.2 Gyr ago or more recently. For most of the favored models from both archetypes, the small planet occurrence-\zmax\ slope, $\tau$, is barely within one standard deviation of the Z23 slope. 


\begin{figure*}
  {\begin{tabular}{@{}cc@{}}
    \includegraphics[width=0.47\textwidth, height=5.5cm]{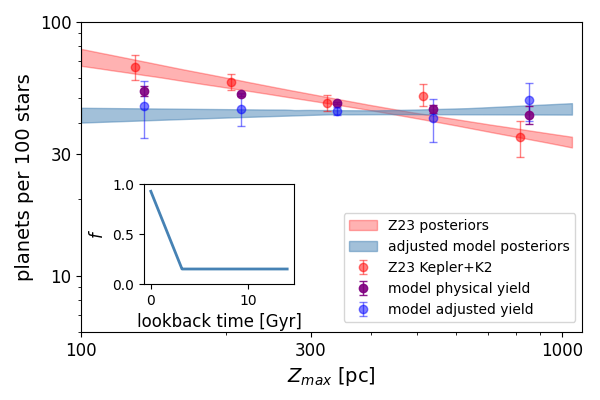} & 
    \includegraphics[width=0.47\textwidth, height=5.5cm]{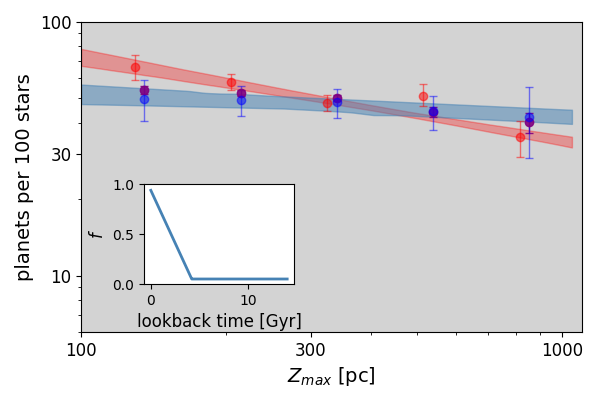} \\ [-7pt]
    \includegraphics[width=0.47\textwidth, height=5.5cm]{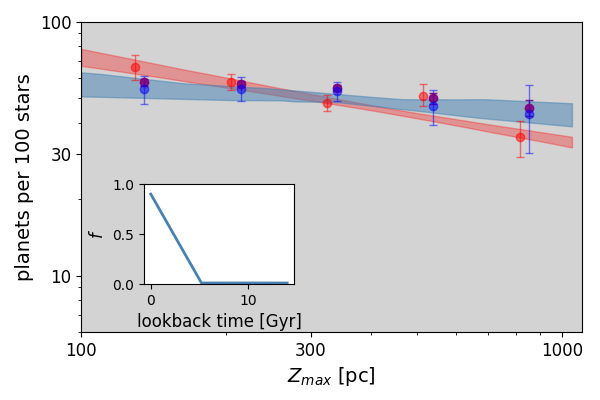} &
    \includegraphics[width=0.47\textwidth, height=5.5cm]{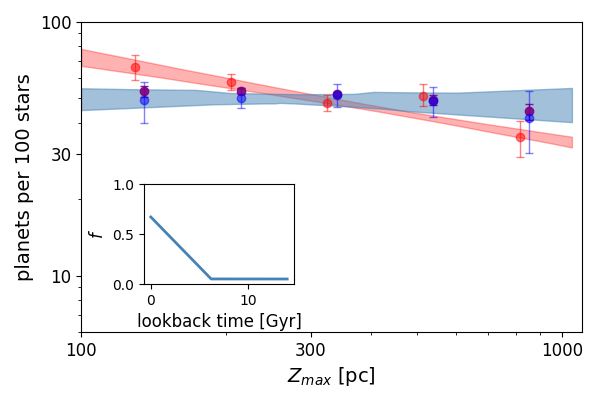} \\ [-7pt]
    \includegraphics[width=0.47\textwidth, height=5.5cm]{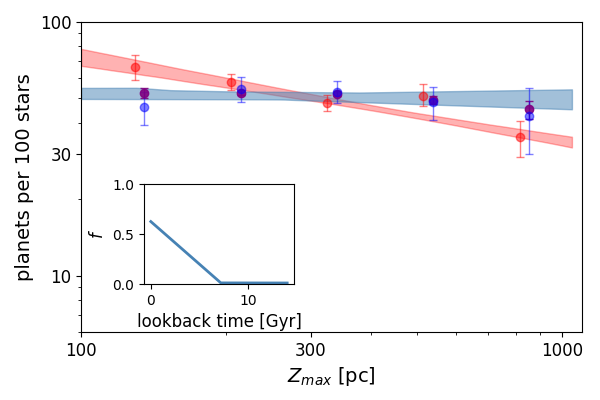} &
    \includegraphics[width=0.47\textwidth, height=5.5cm] {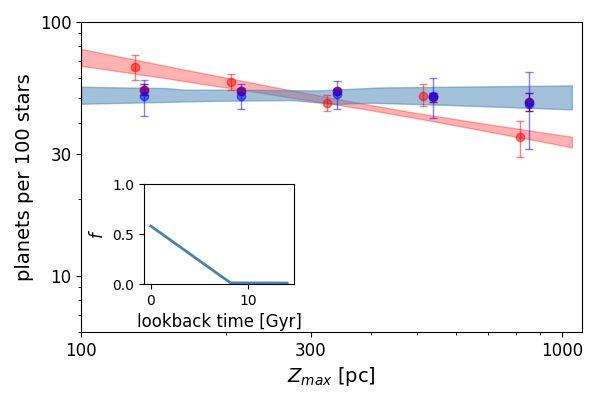}
    \end{tabular}}
    \caption{Planet occurrence versus \zmax\ for six piecewise models with \threshold\ separated by 1 Gyr. Models, shown in insets, prescribe a flat planet host fraction \fone\ until a time \threshold\ at which this fraction gradually rises to some current \ftwo. Favored models are shaded in gray. \textbf{Top left:} \fone=15\%, \ftwo=100\%, \threshold=3.2 Gyr. That is, until $\sim$3.2 Gyr ago, the planet host fraction among Sun-like stars was 15\%; it then increased to a present-day level of 100\%. \textbf{Top right:} \fone=5\%, \ftwo=100\%, \threshold=4.2 Gyr. \textbf{Second row left}: \fone=1\%, \ftwo=95\%, \threshold=5.2 Gyr. \textbf{Second row right}: \fone=5\%, \ftwo=70\%, \threshold=6.2 Gyr. \textbf{Bottom left}: \fone=1\%, \ftwo=65\%, \threshold=7.2 Gyr. \textbf{Bottom right}: \fone=1\%, \ftwo=60\%, \threshold=8.2 Gyr.}
  \label{fig:piecewise}
\end{figure*}

\subsection{Results Using TRILEGAL}
\label{sec:trilegal}
\begin{figure*}
  {\begin{tabular}{@{}cc@{}}
    \includegraphics[width=0.47\textwidth, height=5.25cm]{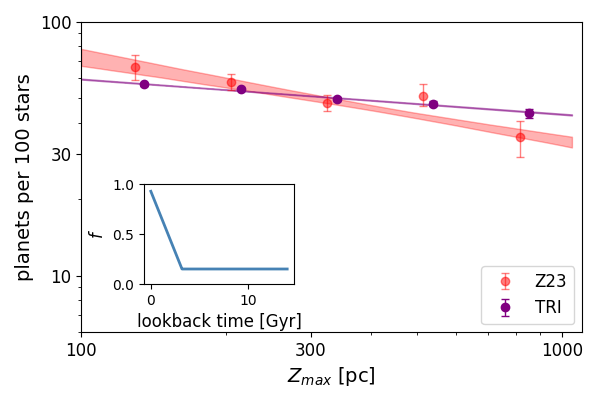} & 
    \includegraphics[width=0.47\textwidth, height=5.25cm]{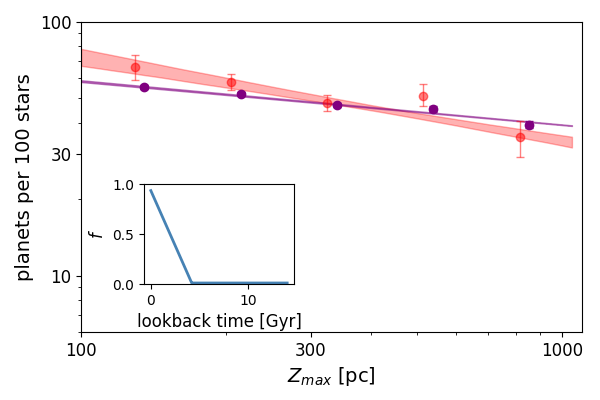} \\ 
    \includegraphics[width=0.47\textwidth, height=5.25cm]{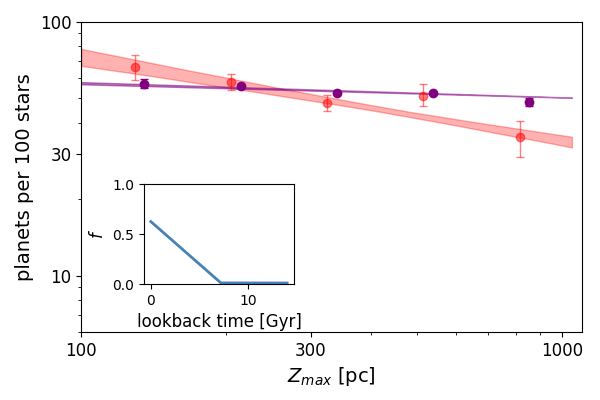} &
    \includegraphics[width=0.47\textwidth, height=5.25cm]{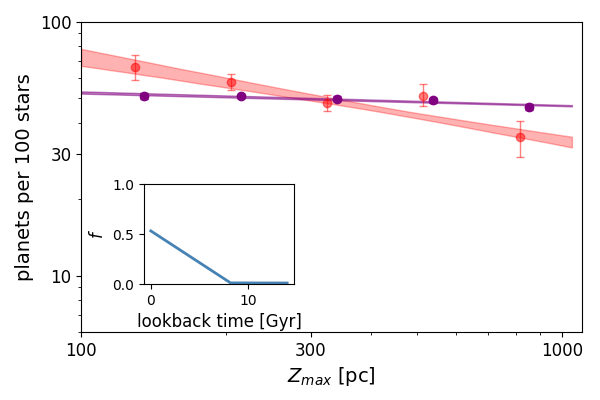} \\
    \includegraphics[width=0.47\textwidth, height=5.25cm]{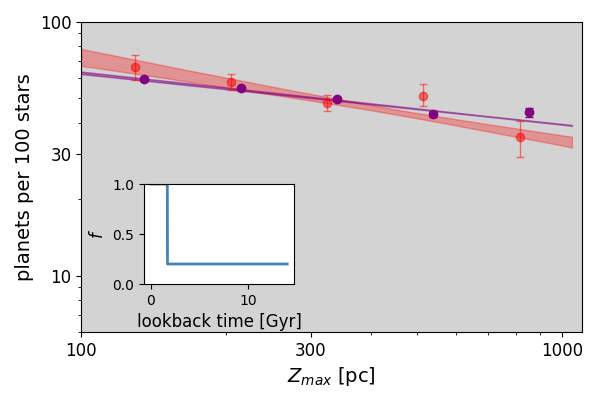} &
    \includegraphics[width=0.47\textwidth, height=5.25cm] {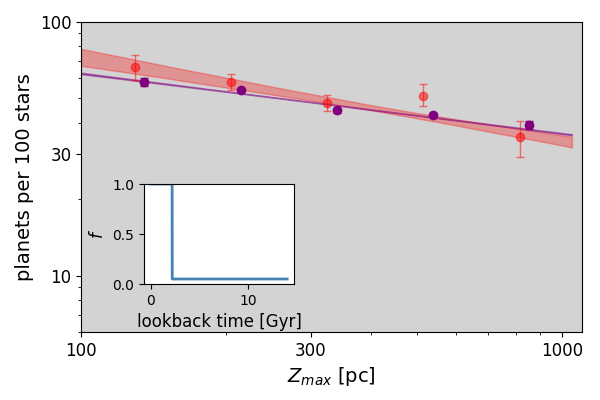} \\
    \includegraphics[width=0.47\textwidth, height=5.25cm]{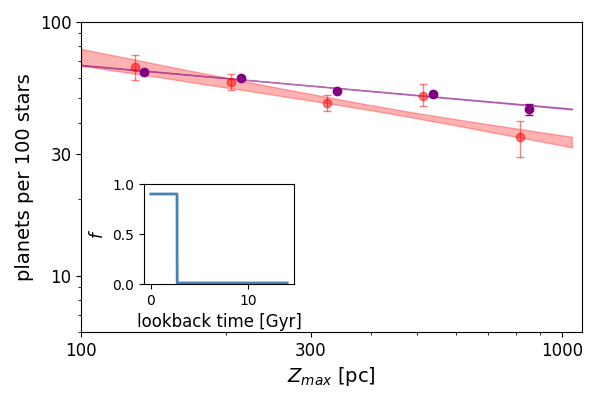} &
    \includegraphics[width=0.47\textwidth, height=5.25cm] {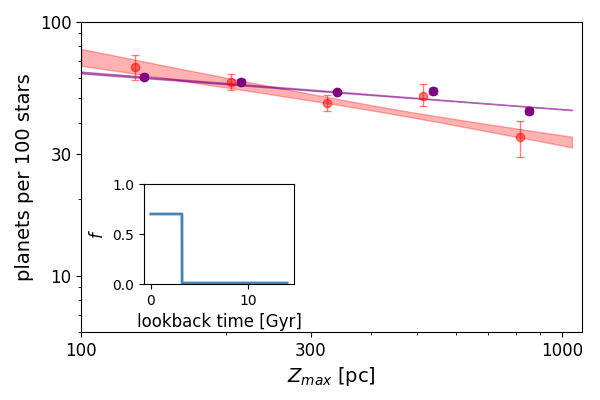}    
    \end{tabular}}
    \caption{Planet occurrence versus \zmax\ for four step and four piecewise models using the TRILEGAL synthetic stellar sample. Favored models are shaded in gray.\textbf{Top left:} Piecewise, \fone=15\%, \ftwo=100\%, \threshold=3.2 Gyr. \textbf{Top right:} Piecewise \fone=1\%, \ftwo=100\%, \threshold=4.2 Gyr. \textbf{Second row left}: Piecewise \fone=1\%, \ftwo=65\%, \threshold=7.2 Gyr. \textbf{Second row right}: Piecewise \fone=1\%, \ftwo=55\%, \threshold=8.2 Gyr. \textbf{Third row left}: Step \fone=20\%, \ftwo=100\%, \threshold=1.7 Gyr. \textbf{Third row right}:  Step \fone=5\%, \ftwo=100\%, \threshold=2.2 Gyr. \textbf{Bottom left}: Step \fone=1\%, \ftwo=90\%, \threshold=2.7 Gyr. \textbf{Bottom right}: Step \fone=1\%, \ftwo=70\%, \threshold=3.2 Gyr.}
  \label{fig:trilegal-models}
\end{figure*}

Without knowing the noise properties of the TRILEGAL stellar samples, we are limited to comparing the true, un-adjusted planet occurrence to Z23. As such, the errorbars reported in Figure \ref{fig:trilegal-models} are much smaller than those in Figures \ref{fig:model-yields-new} and \ref{fig:piecewise}, driven entirely by the age and \zmax\ uncertainties rather than also by the sensitivity functions. Like with HU25-B20-B25, we deploy a battery of step and piecewise functions; however, since the samples differ somewhat in their stellar age distributions, we do not use exactly the same models but instead modify the \fone\ and \ftwo\ for each threshold. Four representative models for each archetype are shown in Figure \ref{fig:trilegal-models}. The piecewise models are anchored by a maximum \ftwo\ at \threshold=3.2 Gyr that is unable to achieve a steep enough $\tau$ and a minimum \fone\ at \threshold=8.2 Gyr that is similarly too shallow. The step models are anchored by a minimum \fone\ at a relatively recent \threshold=3.2 Gyr that is too shallow, but unlike with HU25-B20-B25, we are able here to identify models prescribing recent step increases in \f\ that fit more closely to the Z23 result. Step function \{\fone=20\%, \ftwo=100\%, \threshold=1.7 Gyr\} yielded a $\tau$ of -0.20$\pm$0.02 and $\eta$ of 0.46$\pm$0.01, and step function \{\fone=5\%, \ftwo=100\%, \threshold=2.2 Gyr\} yielded a $\tau$ of -0.24$\pm$0.02 and $\eta$ of 0.43$\pm$0.00. Despite the better match, these two models are tuned to the theoretical maximum \ftwo=100\%.

\subsection{Results Using Only Kepler}
\label{sec:kepler-only}
Thus far, we have sought to match the Z23 joint \textit{Kepler}-K2 occurrence rate, but our study is limited by a lack of K2 stars with age measurements. We now consider what the \textit{Kepler} planet occurrence-\zmax\ relation would look like, and whether an instantaneous or gradual rise in \f\ would be able to reproduce this trend. We prepared the stellar and planetary sample as described in Z23, but only for the \textit{Kepler} field, and propagate the observed planet occurrence-\zmax\ trend through the same completeness pipeline in Z23. Without \zmax-disaggregated sensitivity maps, we simply bin all planets by in 9x9 radius and period space regardless of \zmax\ and multiply by the reliability map and divide by the sensitivity map from \cite{thompson_planetary_2018}. We divide this by the same geometric transit probability map from the joint HU25-B20-B25 sample, and then sum the adjusted planets per \zmax\ bin to get the \textit{Kepler} planet occurrence per 100 stars as a function of \zmax: \{49.1$\pm$3.1, 43.4$\pm$1.7, 36.5$\pm$1.5, 37.4$\pm$2.4, 30.2$\pm$3.8\}. We fit this using the same power law from Z23 and get a best-fit $\tau$ of -0.25$\pm$0.05 and $\eta$ of 0.35$\pm$0.01. While the \textit{Kepler} occurrence is understandably less than the Z23 result (by $\sim$0.10, which suggests that that is the K2 small planet occurrence around FGK dwarfs), it is notable that the planet occurrence-\zmax\ trend slope is essentially the same. This suggests that the observed trend from Z23 should hold across the \textit{Kepler} and K2 fields. 

Using this as our new ``ground truth", we attempt to model step and piecewise increases in \f\ over time as before. We find a relatively good match in the step function {\fone=20\%, \ftwo=100\%, \threshold=2.2 Gyr\} (depicted in the top panel of Figure \ref{fig:kepler-only-model}), which produced a $\tau$ of -0.20$\pm$0.07 and a $\eta$ of 0.34$\pm$0.01. This model has the same \threshold\ and \ftwo\ as one of the successful models for HU25-B20-B25, but it has a higher \fone, likely due to differences in the age distribution between the joint and \textit{Kepler}-only samples. The latter contains more older stars, requiring a boost to either \fone\ or \ftwo\ to maintain the correct occurrence normalization. The age histograms  are shown in the bottom panel of Figure \ref{fig:kepler-only-model}. This difference in age distributions may also be responsible for the slight bump in planet occurrence at the 300 pc \zmax\ bin. This deviation from a monotonically decreasing trend is persistent across model thresholds and archetypes. 

\begin{figure}
\includegraphics[width=.45\textwidth, height=5.25cm]{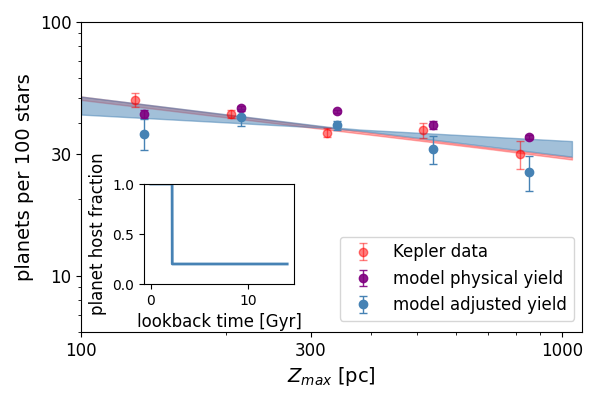} \\
\includegraphics[width=.45\textwidth, height=5.25cm]{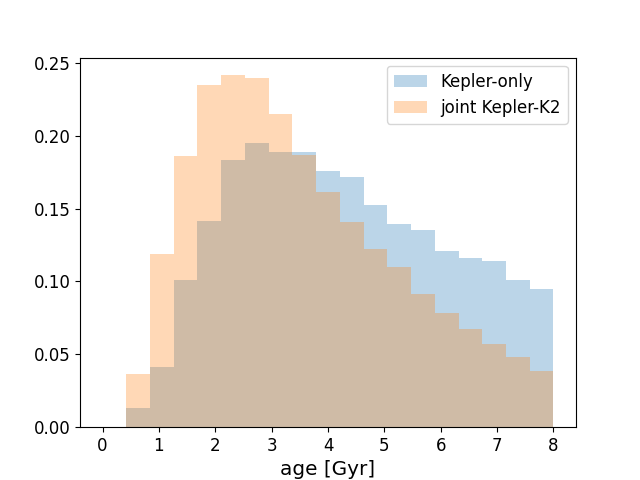} 
\caption{\textbf{Top:} Planet occurrence versus \zmax\ for a fiducial step function model applied to just the \textit{Kepler} sample. The model used here is \fone=5\%, \ftwo=80\%, \threshold=9.5 Gyr. Using only the \textit{Kepler} field, it is clear we require a broken power law to better model the occurrence-\zmax\ trend. \textbf{Bottom:} Stellar age distribution for the \textit{Kepler}-only versus the joint \textit{Kepler}-K2 sample used in the bulk of this study. The former is shifted older and flatter, which may explain the non-monotonic trend in small planet occurrence with \zmax\ for this sample.} 
\label{fig:kepler-only-model}
\end{figure}

\section{Discussion}
\label{sec:discussion}
We have demonstrated in the preceding sections that an increase in the planet host fraction over the last 8 Gyr -- however gradual -- is insufficient to reproduce the steepness of the observed decrease in small planet occurrence with increasing maximum oscillation amplitude from the Galactic midplane. We synthesized stellar and planet populations following a battery of models in which the planet host fraction, \f, increased either instantaneously or over Gyr timescales, with the increase (or onset of the increase) occurring at various times in the Milky Way's past. Of these models, we ruled out all but the following: a step increase 2.2 Gyr ago from \fone\ = 10\% to \ftwo\ = 100\%, a step increase 2.7 Gyr ago from \fone\ = 1\% to \ftwo\ = 95\%, a step increase 3.2 Gyr ago from \fone\ = 1\% to \ftwo\ = 75\%, a gradual increase starting 4.2 Gyr ago from \fone\ = 5\% to \ftwo\ = 100\%, and a gradual increase starting 5.2 Gyr ago from \fone\ = 1\% to \ftwo\ = 95\%. These models are not ruled out because their best-fit $\tau$ and $\eta$ are within a standard deviation from the Z23 combined and re-fit Super-Earth and Sub-Neptune $\tau$ and $\eta$. However, these models' best-fit $\tau$ lie on the edge of the acceptable $\tau$ envelope of $\tau$ = 0.28$\pm$0.08. Moreover, for all of these models, \fone\ or \ftwo\ was pushed to an unrealistic minimum or maximum, respectively. In two cases -- the step increase 2.7 Gyr ago and the gradual increase beginning 5.2 Gyr ago -- we nearly pushed both \fone\ and \ftwo\ to their extrema. 

It is informative that almost none of our models applied to the HU25-B20-B25 sample yield a relation that is steeper than the Z23 slope. The Z23 stellar sample did not have stellar ages, but it is possible at least to compare their \zmax\ distribution, which peaks between 288 and 408 pc, to that of our joint sample, which peaks at 263 pc. While we constructed our sample in much the same way as Z23, our requirement that stars in our sample must have age measurements means that our samples are not identical. Most significantly, our sample's \zmax\ distribution is bunched around lower \zmax, leading to proportionally larger errorbars for the last two Galactic height bins and reduced sensitivity to changes in planet occurrence over this \zmax\ region. 

The TRILEGAL sample likewise has a lower instantaneous height peak and smaller MAD than the \textit{Kepler} and K2 \zmax\ in Z23. However, it has much more precise stellar ages than the HU25-B20-B25 sample. We demonstrated in Figure \ref{fig:trilegal-models} that a late-time instantaneous rise in \f\ can match the Z23 result. While this shows that it is possible in principle for an increase in \f\ to produce a strongly downward sloping planet occurrence with increasing \zmax, the requirement among the best-fit models that either \fone\ = 1\% or \ftwo\ = 100\% suggests that \f\ cannot single-handedly drive the overall small planet occurrence-\zmax\ relation. 

Future work could be more successful at matching Z23 by adding complexity to our model in two ways. First, Z23 conducted their analysis on Super-Earths and Sub-Neptunes separately, while our models were agnostic about this distinction. In Z23, it was the Sub-Neptunes that drove the steepness of the planet occurrence-\zmax\ relation, whereas the Super-Earth occurrence was flat until 500 pc. Therefore, allowing for different \f\ for Super-Earths and Sub-Neptunes can build in the flexibility required to match the Z23 trend. Second, while \texttt{psps} allows users to modulate the fraction of compact multis, in this study we have left the so-called ``intact" fraction relatively constant at 18$\pm$10\% \citep{lam_ages_2024}. Allowing this to be a free parameter in addition to \f\ could provide the means to produce greater differences in planet occurrence for stars of different ages. Since the issue with most of our models is that they are too shallow, we posit that swapping an increase in \f\ with an increase in the intact fraction could result in models that better fit the Z23 result. 

Another potential systematic cause of the diluted occurrence-\zmax\ trend in our models is the weak age-\zmax\ relation in our joint \textit{Kepler}-K2 sample. This relation was measured to be 4 Gyr/kpc using asteroseismic ages \citep{casagrande_measuring_2016}. ESA's upcoming PLAnetary Transits and Oscillations of stars (PLATO) mission will measure precise solar-like oscillations for 245,000 FGK dwarf stars, enabling much more precise ages for planet hosts on a demographically relevant scale \citep{rauer_plato_2014, montalto_all-sky_2021, boettner_exoplanets_2024, goupil_predicted_2024}. With a large, uniform sample of relatively high-precision ages, it will be possible to re-examine age-based models of Galactic-scale exoplanet demographics. 

While the models we have failed to rule out are not very plausible, we would be remiss not to briefly explore their implications on planet formation in the Milky Way's history, if only to present a framework for future work. Among all the step and piecewise models for time-varying planet occurrence that we applied to the HU25-B20-B25 and TRILEGAL joint \textit{Kepler}-K2 stellar samples, all models that matched Z23 were step functions and had \threshold\ no earlier than 3.2 Gyr ago. If true, this is suggestive of a single event rather than a series of events (or a more gradual process such as Galactic Chemical Evolution) that is overwhelmingly responsible for the strong correlation between planet occurrence and \zmax. Such a process would have had to be relevant for not only the \textit{Kepler} field but also the K2 fields. In the following paragraphs, we discuss whether certain Galactic scale events in the Milky Way's past could have coincided with one of these late-time boosts to \f. These events include major and minor mergers, close passages of satellite galaxies, and global gravitational instabilities or external infall, which could drive fluctuations in the interstellar medium (ISM).

Galaxy major mergers are thought to be one of the main factors triggering star formation in galaxies and are predicted to have clear effects on their chemical evolution \citep{tinsley_evolution_1980, tissera_double_2002, ellison_galaxy_2013, ruiz-lara_recurrent_2020}. The timing of these dynamical events is a potential clue, but we must also consider the nature and scale of the actual processes that might affect planet occurrence. We expect mergers to affect the planet occurrence rate chemo-kinematically by adding gas of lower and heterogeneous metallicity to the planet forming budget and stirring up the interstellar medium (ISM). Mergers could also compress the ISM during passage and thus stimulate not only additional star formation but also persistent spiral structure \citep{2013ApJ...766...34D}. On the other hand, close passages might contribute to ISM turbulence without modifying its chemistry. These processes could also each have competing effects for planet formation. For example, \citet{lu_turning_2022} showed that the injection of metal-poor gas from satellite infall could have triggered the formation of metal-poor stars as recently as $\sim$3 Gyr ago, which would affect planet occurrence among younger stars.

While not all of these mechanisms directly affect planet formation, they can impact (1) the formation conditions for protoplanetary disks and (2) the kinematic signatures or even survival of mature planetary systems. To that end, we can ask, based on our timing prescriptions in this paper: \textit{does the timing line up for when we need a planet-moderating effect to occur}? Sources in the literature point to a number of potential significant dynamical interactions that our Galaxy has had with external bodies over the course of its history: the Gaia-Sausage-Enceladus (GSE) dwarf galaxy merger 8-10 Gyr ago \citep{helmi_merger_2018, belokurov_co-formation_2018}; the Sagittarius dwarf spheroidal (Sgr dSph) galaxy first, second, and third passages 5.7, 1.9, and 1.0 Gyr ago, respectively \citep{ruiz-lara_recurrent_2020}; and the Virgo Radial Merger (VRM) 2.7 Gyr ago \citep{donlon_virgo_2019, donlon_milky_2020, donlon_debris_2024}, sometimes called the Virgo Overdensity, \citep[VOD;][]{2001ApJ...554L..33V}. Additionally, satellite infall of dwarf galaxies spans the Milky Way's history, but we are primarily interested in ones large and close enough to significantly shift the planet formation or evolution conditions in the Solar neighborhood. The favorable window includes the VRM merger event \citep{donlon_virgo_2019, donlon_milky_2020, 2022ApJ...932L..16D, donlon_debris_2024}, a putative radial dwarf galaxy merger approximately 2.7~Gyr ago that produced debris in the Solar neighborhood.  It is also probable that the Milky Way disk has experienced multiple gas-rich radial mergers  \citep{2020MNRAS.498.2472K, 2021MNRAS.500.1385H, 2022ApJ...932L..16D, 2023ApJ...944..169D} similar to the VRM that could have stimulated episodes of enhanced star formation. The Sagittarius dwarf galaxy's second passage occurred \citep{ruiz-lara_recurrent_2020} 1.9~Gyr ago and coincided with a narrow period of enhanced star formation. However, given that Sagittarius' closest passage was at ${\gtrsim}15$~kpc from the Galactic center, it is unclear how much this would have impacted planet formation in the Solar Neighborhood. 


Our step function models' \threshold\ parameters mark the timing of some of these events: \threshold=1.7~Gyr corresponds to 200~Myr after the second passage of the Sagittarius dwarf galaxy; \threshold=2.2~Gyr corresponds to 500~Myr after the VRM; \threshold=6.2~Gyr corresponds to shortly before the first passage of Sgr dSph; and \threshold=8.2~Gyr lies toward the end of the window of the Gaia-Enceladus merger. \threshold=4.2~Gyr marks a fiducial waypoint in the Milky Way's dynamical history. It is difficult to interpret more recent \threshold\ because of our sample's large isochrone age uncertainties, while probing \threshold\ at cosmological times earlier than $\sim$5.5~Gyr would require extrapolating beyond the oldest stars in our sample. Using strictly timing arguments, therefore, our analysis can rule out the imprint of the GSE merger and the first passage of Sgr dSph on planet occurrence in our Galactic neighborhood.  


Mergers or close passages offer a potentially illuminating framework here in another way, which is to mediate the apparent link between planet occurrence and stellar kinematics \citep{bashi_exoplanets_2022, kruijssen_not_2021, yang_planets_2023}. Infalls themselves could have dynamically heated stars in the path of their trajectories, puffing up the orbits of older stars to greater scale heights. Similarly, stellar radial migration \citep{2020ApJ...896...15F,2019ApJ...882..111D, 2015MNRAS.447.3576D,2010ApJ...722..112M,2002MNRAS.336..785S} could have moved stars from the metal-rich center outward, and from the metal-poor periphery inward, where there is evidence that the Sun itself is an interloper from the inner disk \citep{2024MNRAS.535..392L,1996A&A...314..438W}. On the other hand, older populations may very well have formed in a thick, turbulent disk \citep[e.g.][]{2025arXiv250201895B,McCluskey2024,2020MNRAS.492.4716B,2009ApJ...707L...1B}. There is an important relationship between \zmax, the subject of Z23, and vertical action, $J_{z}$: both are dynamical quantities that capture the amplitude of a star’s oscillation above and below the Galactic midplane and generally increase with age \citep{beane_implications_2019, spitoni_disc_2022, sagear_zoomies_2024, price-whelan_data-driven_2025}. Action-angle coordinates, including $J_{z}$, are powerful tools for tracing a star’s long-term kinematic history due to dynamical disturbances \citep{sanders_review_2016}. Famously, the spiral substructure of the ``Antoja snail" \citep{antoja_dynamically_2018} is encoded in action-angle coordinates and possibly attributable to the passage of Sagittarius referenced above \citep{laporte_footprints_2019} -- in this sense, an occurrence dependent upon \zmax\ may be interpretable in the context of galactic disturbances.

For now, it is not clear whether information from galactic processes (e.g. mergers, satellite in-fall) can reliably propagate down to local processes such as turbulence, or whether the local ISM turbulence at planet formation sites is instead causally disconnected from Galactic chemical evolution (GCE) and kinematics. \cite{winter_planet_2024} suggested that ISM turbulence could prolong protoplanetary disk lifetimes among 20-70\% of systems, which could affect the final accounting of mature planets. A low level of turbulence can enable local vortices at planet formation sites to efficiently trap and concentrate dust, lowering the threshold for planetesimal formation from near solar-like metallicity down to 0.08 Z\textsubscript{Solar} for Mars-like planets \citet{eriksson_planets_2025}, although higher turbulence levels may require higher metallicity thresholds for planet formation. Observational evidence was shown by \citet{marchi_protoplanetary_2024}, who found that protoplanetary disks around metal-poor Sun-like stars are longer-lived than previously thought. If the average ISM turbulence were to be more favorable among younger metal-poor stars than older ones, this could contribute to the observed trend in Z23. 

We have so far focused upon some process \textit{enhancing} recent star and planet formation relative to some former baseline. A seeming recent rise in recent planet occurrence could also be produced by a process progressively carving away at older planetary systems. Such a process would act by \textit{diminishing} planet occurrence preferentially around older stars, perhaps in a way where the cumulative probability of destruction or ejection accrues with time. The above scenarios that we have touched upon so far invoke galactic dynamical history as a way to drive bursts in star (and potentially planet) formation, rather than directly acting on the planets \textit{after} formation. Stellar flybys have been suggested as a possible direct mechanism for planet loss and the altering of system architectures \citep{zakamska_excitation_2004, rodet_correlation_2021}. In concert with additional secular effects like Kozai-Lidov \citep{naoz_eccentric_2016}, close stellar flybys in dense cluster environments can sometimes lead to Hot Jupiters and ultra-cold Saturns, although they more often lead to planet ejections \citep{wang_hot_2020}. \citet{charalambous_breaking_2025} show through N-body simulations that stellar flybys as far away as 1000~AU can disrupt low-order resonant chains, although smaller planets are relatively more resilient to this effect. Along with \citet{schoettler_effect_2024}, they also demonstrate cases in which a delayed disruption of the architecture occurs, over 100~Myr timescales. It is also possible that planet formation was suppressed earlier on from the radiation environment during formation. \citet{hallatt_formation_2025} showed that the protoplanetary disks of thick disk stars, which formed closer to cosmic noon, experienced $\sim$7 orders of magnitude more background radiation than Solar neighborhood stars, limiting their lifetimes to 0.2-0.5 Myr. 

The menu of culprits listed above that could affect planet formation and survivability is long but can be divided into varying operative timescales. If we adopt the premise that a late-time apparent increase in planet formation has occurred, the balance of the processes listed above must continually result in an increasingly net gain of planets over time. Sudden jumps in planet occurrence at earlier cosmological times cannot explain the Z23 result. This means that present-day exoplanet demographics were either not set by earlier Galactic-scale events (e.g. the Gaia-Enceladus merger), or whatever imprints were left by those events have now been erased. Determining how this occurs -- perhaps through some combination of deleterious effects that become ameliorated over time (e.g. elevated radiation environment in high-traffic dense birth clusters) and increasingly fertile grounds for planet formation (e.g. the average local ISM turbulence becoming more favorable among more recently-formed metal-poor stars) -- would require synthesizing observational and simulation studies of star and planet formation sites encompassing the Solar neighborhood and spanning a $\pm$1 kpc column on either side of the Galactic midplane.

\section{Conclusion}
\label{sec:conclusion}

The impact of the galactic environment on exoplanet demographics has been the subject of much recent study. \citet{zink_scaling_2023} showed that the metallicity gradient expected from Galactic chemical evolution is insufficient to explain a decreasing short-period planet occurrence with increasing galactic height, as observed by the \textit{Kepler} and K2 missions. In this work, we forward modeled two prescriptions for increasing recent planet occurrence, as well as a flat (control) model, producing observed synthetic populations of planetary systems using the software package \texttt{psps}.  By assigning planet populations, tuned to reflect different ``burst" times in planet occurrence, to a synthetic \textit{Kepler} stellar sample, we show that the timing and nature of that burst produce a range of slopes in \zmax\ versus planet occurrence space. 

Generally speaking, we find that an increase in the small planet host fraction among FGK dwarfs is insufficient to reproduce the steepness of the downward trend in planet occurrence with increasing \zmax. There are some models -- exclusively step increases (rather than more gradual increases in \f) -- at recent Galactic times that can match the Z23 result to within one standard deviation, but these models are tuned to implausibly high \ftwo\ or low \fone, and they are not strong matches. Part of this may be attributable to our lack of precise stellar ages (or, indeed, ages of any quality for K2 stars), but even an idealized 1 Gyr age uncertainty from the TRILEGAL sample does not appreciably change the intrinsic planet occurrence yields of the models explored in this work. As future work, perhaps tuning the intact fraction rather than the planet host fraction will provide a better match to Z23. In this study, we also independently derived the \textit{Kepler} small planet occurrence trend with \zmax\ and find that it is relatively easier for a step increase in \f\ to match this slope. 

We have focused in this manuscript about the \textit{timing} of the effect required, rather than a particular mechanism. We center our discussion upon the ways that galactic processes could potentially drive planet formation on the right timescales, but as yet, we cannot distinguish between a process that enhances recent planet formation, versus one that mimics the same trend by progressively carving away at older planetary systems. A satisfactory theory of the role of Galactic-scale events in planet formation and survivability requires making the connection between the spatial scales of these two seemingly disparate processes. Specifically, it is important to pose one or more plausible physical mechanisms for directly affecting either the planet formation stage or the ability of mature planetary systems to keep hold of their planets. There are major conceptual challenges in connecting the spatially and temporally extended processes relevant to the Galaxy on Gyr timescales to the smaller-scale physics of planet assembly and evolution. However, the canonical picture of local metal enrichment as the only galactic-planetary connective tissue cannot fully explain planet demographics in the \textit{Gaia} era. The puzzles of why thin/thick disk planet populations differ \citep{adibekyan_chemical_2012, adibekyan_compositional_2021, bashi_exoplanets_2022, hallatt_formation_2025, boettner_exoplanets_2024}, whether and how common kinematics drive patterns in planet demographics \citep{winter_stellar_2020, rampalli_disentangling_2025, yang_planets_2023, mustill_hot_2022, kruijssen_not_2021, kruijssen_bridging_2020, blaylock-squibbs_no_2024}, and whether and how planetary systems evolve over Gyr in isolation or in response to external perturbations \citep{kaib_planetary_2013, pu_spacing_2015, sayeed_exoplanet_2025, chen_planets_2021, miyazaki_evidence_2023, rodet_correlation_2021} merit a larger-scale contextual discussion. Continued observational and simulation studies connecting GCE and Galactic dynamics to giant molecular cloud (GMC) and ISM dynamics, and then connecting GMCs and the ISM to protoplanetary disks, will enable stronger statements about the role of the Galactic context in exoplanet demographics. 

\section{Acknowledgments}
\label{sec:acknowledgments}
We wish to thank Luke Bouma, Jon Zink, Pat Tamburo, Desika Narayanan, Jamie Tayar, Adrian Price-Whelan, Carrie Filion, George Privon, Quadry Chance, Natalia Guerrero, Jason Dittmann, William Schap III, and the UF Astronomy Department Stars \& Planets Journal Club for their helpful comments and suggestions. This work has made use of data from the European Space Agency (ESA) mission {\it Gaia} (\url{https://www.cosmos.esa.int/gaia}), processed by the {\it Gaia} Data Processing and Analysis Consortium (DPAC, \url{https://www.cosmos.esa.int/web/gaia/dpac/consortium}). Funding for the DPAC has been provided by national institutions, in particular the institutions participating in the {\it Gaia} Multilateral Agreement. This research has made use of the NASA Exoplanet Archive \citep{christiansen_nasa_2025}, which is operated by the California Institute of Technology, under contract with the National Aeronautics and Space Administration under the Exoplanet Exploration Program. This material is based upon work supported in part by the National Science Foundation GRFP under Grant No. 1842473. 

KJD acknowledges support from the Heising Simons Foundation grant \# 2022-3927. She also respectfully acknowledge that the University of Arizona is home to the O'odham and the Yaqui. She respects and honors the ancestral caretakers of the land, from time immemorial until now, and into the future.

We acknowledge that IPAC/Caltech resides on the traditional, ancestral and unceded territory of the Gabrielino/Tongva peoples, the original caretakers of this land. We respectfully recognize the Gabrielino/Tongva peoples who still reside in Tovaangar (the Los Angeles basin and South Channel Islands). We also acknowledge that the main campus of the University of Florida is located on the ancestral territory of the Potano and of the Seminole peoples. The Potano, of Timucua affiliation, lived here in the Alachua region from before European arrival until the destruction of their towns in the early 1700s. The Seminole, also known as the Alachua Seminole, established towns here shortly after but were forced from the land as a result of a series of wars with the United States known as the Seminole Wars. We, the authors, acknowledge our obligation to honor the past, present, and future Native residents and cultures of California and Florida.

\facility{Kepler, Gaia}


\bibliography{main}
\bibliographystyle{aasjournal}

\end{document}